\documentclass[twocolumn]{aastex63}
\usepackage{amsmath}
\usepackage{here}
\usepackage{graphics}

\newcommand{\rhog}{\rho_{\mathrm{g}}}
\newcommand{\rhod}{\rho_{\mathrm{d}}}
\newcommand{\rhogup}{\rho_{\mathrm{g},0}}
\newcommand{\rhodup}{\rho_{\mathrm{d},0}}
\newcommand{\gas}{\mathrm{g}}
\newcommand{\dst}{\mathrm{d}}

\newcommand{\sigmag}{\Sigma_{\gas}}

\newcommand{\sigmad}{\Sigma_{\dst}}

\newcommand{\tstop}{t_{\mathrm{stop}}}

\newcommand{\taus}{\tau_{\mathrm{s}}}

\newcommand{\cs}{c_{\mathrm{s}}}

\newcommand{\hd}{H_{\mathrm{d}}}

\newcommand{\qthree}{Q_{\mathrm{3D,\gas}}}
\newcommand{\zs}{Z_{\mathrm{s}}}
\newcommand{\zsd}{Z_{\mathrm{s,d}}}
\newcommand{\fdiffx}{F_{\mathrm{diff},x}}

\def\msun{M_{\odot}}
\def\merth{M_{\oplus}}

\received{--}
\accepted{--}
\submitjournal{ApJ}

\shorttitle{Stratified Secular GI}
\shortauthors{Tominaga et al.}
\begin{document}

\title{On Secular Gravitational Instability in Vertically Stratified Disks}

\correspondingauthor{Ryosuke T. Tominaga}
\email{ryosuke.tominaga@riken.jp}
\author[0000-0002-8596-3505]{Ryosuke T. Tominaga}
\affiliation{RIKEN Cluster for Pioneering Research, 2-1 Hirosawa, Wako, Saitama 351-0198, Japan}

\author[0000-0003-4366-6518]{Shu-ichiro Inutsuka}
\affiliation{Department of Physics, Nagoya University, Nagoya, Aichi 464-8692, Japan}

\author[0000-0003-3038-364X]{Sanemichi Z. Takahashi}
\affiliation{National Astronomical Observatory of Japan, Osawa, Mitaka, Tokyo 181-8588, Japan}
\affiliation{Graduate School of Science and Engineering, Kagoshima University, Kagoshima 890-0065, Japan}
%
%\author{Ryosuke T. Tominaga}
%\affiliation{RIKEN Cluster for Pioneering Research, 2-1 Hirosawa, Wako, Saitama 351-0198, Japan}
%\affiliation{Department of Physics, Nagoya University, Nagoya, Aichi 464-8692, Japan}
%    
%\author{Shu-ichiro Inutsuka}
%\affiliation{Department of Physics, Nagoya University, Nagoya, Aichi 464-8692, Japan}
%
%\author{Hiroshi Kobayashi}
%\affiliation{Department of Physics, Nagoya University, Nagoya, Aichi 464-8692, Japan}



\begin{abstract}
Secular gravitational instability (GI) is one  promising mechanism for explaining planetesimal formation. The previous studies on secular GI utilized a razor-thin disk model and derived the growth condition in terms of the vertically integrated physical values such as dust-to-gas surface density ratio. However, in weakly turbulent disks where secular GI can operate, a dust disk can be orders of magnitude thinner than a gas disk, and analyses treating the vertical structures are necessary to clarify the interplay of the midplane dust motion and the upper gas motion. In this work, we perform vertically global linear analyses of secular GI with the vertical domain size of a few gas scale heights. We find that dust grains accumulate radially around the midplane while gas circulates over the whole vertical region. We obtain well-converged growth rates when the outer gas boundary is above two gas scale heights. The growth rates are underestimated if we assume the upper gas to be steady and regard it just as the source of external pressure to the dusty lower layer. Therefore, treating the upper gas motion is important even when the dust disk is much thinner than the gas disk. Conducting a parameter survey, we represent the growth condition in terms of the Toomre's $Q$ value for dust and dust-to-gas surface density ratio. The critical dust disk mass for secular GI is $\sim10^{-4}$ stellar mass for the dust-to-gas surface density ratio of 0.01, the Stokes number of 0.1, and the radial dust diffusivity of $10^{-4}\cs H$, where $\cs$ is the gas sound speed and $H$ is the gas scale height.
\end{abstract}



\section{Introduction}\label{sec:intro}
Formation of planetesimals from dust grains has been widely investigated. It has been a long-standing issue that the growth of compact dust grains toward planetesimals is prevented by some processes including fast radial drift and collisional fragmentation \citep[e.g.,][]{Weidenschilling1977,Weidenschilling1993,Blum2000}. One possible process to promote planetesimal formation is dust clumping due to disk instabilities. Streaming instability is regarded as the most promising process and has been extensively studied \citep[e.g.,][]{YoudinGoodman2005,Youdin2007,Johansen2007,Johansen2007nature,Bai2010a,Bai2010b,Bai2010c,Yang2017,Abod2019,Krapp2019,Chen2020,Umurhan2020,Paardekooper2020,Paardekooper2021,McNally2021,Zhu2021,Yang2021,Li2021,Carrera2021,Carrera2022}. Streaming instability develops at a spatial scale much shorter than the gas scale height \cite[e.g., see][]{Youdin2007} and creates dust clumps that can develop toward planetesimals via gravitational collapse \cite[e.g.,][]{Johansen2007nature,Simon2016,Gerbig2020}. Gravitational collapse of dust clumps is expected to explain the observed frequency of prograde binaries among the Kuiper Belt objects \citep[e.g.,][]{Nesvorny2019,Visser2022}.

Recently, some studies found that streaming instability is significantly stabilized by dust diffusion due to gas turbulence \citep[][]{Umurhan2020,Chen2020,Gole2020}. \citet{McNally2021} performed a linear analysis not only with dust diffusion but also with a dust size distribution. They found that even weak turbulence whose dimensionless strength $\alpha$ \citep[][]{Shakura1973} is $\sim10^{-5}-10^{-3}$ stabilizes streaming instability unless the dust-to-gas ratio is sufficiently higher than in inviscid cases (see Figure 8 therein). Since such low $\alpha$ values are indicated from recent disk observations \citep[e.g.,][]{Flaherty2015,Flaherty2017,Pinte2016,Villenave2022}, streaming instability might be inefficient to form planetesimals. Therefore, other dust-clumping mechanisms are important to reveal planetesimal formation: such as coagulation instability \citep[][]{Tominaga2021,Tominaga2022a,Tominaga2022b} and azimuthal drift-driven streaming instability \citep{Lin2022,Hsu2022}.

In this work, we focus on secular gravitational instability (GI), which is another mechanism potentially explaining planetesimal formation \citep[e.g.,][]{Ward2000,Youdin2005a,Youdin2005b,Youdin2011,Shariff2011,Takeuchi_Ida2012,Michikoshi2012,Takahashi2014,Tominaga2018,Tominaga2019,Tominaga2020,Pierens2021}. Secular GI is one kind of gas-drag-driven instability. In contrast to the classical GI \citep[e.g.,][]{Goldreich1965a,Sekiya1983} and drag-mediated two-fluid GI \citep[e.g.,][]{Coradini1981,Longarini2022}, secular GI operates even in GI-stable disks and concentrates dust grains into rings. Its nonlinear development leads to not only gravitational collapse of the dust ring but also azimuthal fragmentation via Rossby wave instability \citep[e.g.,][]{Lovelace1999,Li2000}, and is expected to cause planetesimal formation \citep[][]{Tominaga2020,Pierens2021}. As demonstrated by \citet{Takahashi2022}, the gravitational fragmentation of the dust ring creates prograde dust clumps. This is expected to explain the origin of the prograde binaries of the Kuiper Belt objects as in the case of streaming instability. 

The previous studies on secular GI utilized a razor-thin disk model with the two-fluid formalism of dust and gas. In the model, the drag-driven momentum transport between dust and gas is averaged within the full vertical extent of a disk. Its efficiency is thus assumed to be proportional to the dust-to-gas surface density ratio. However, the aerodynamical dust-gas interaction should operate only within the dust sublayer, which can be substantially thin in a weakly turbulent gas disk. This gap in the vertical scales of gas and dust makes the previous growth condition of secular GI ambiguous as pointed out by \citet{Latter2017}. For example, in the thin disk model where the momentum transport is averaged within a gas disk, we used Toomre's $Q$ value of a gas disk to discuss the stability. This value is inversely proportional to the gas surface density $\Sigma_{\gas}$. One may use the effective $Q$ value by reducing $\Sigma_{\gas}$ to the gas surface density within the dust sublayer where the aerodynamical interaction is efficient. \citet{Latter2017} found that the previous growth condition is difficult to be satisfied in GI-stable gas disks if one adopts this effective $Q$ value by ignoring gas above the dust sublayer \citep[see also][]{Lesur2022}. This contradicts the original advantage of secular GI over the classical GI. 

To address the above issue, we conduct three-dimensional linear analyses with the vertical structures and vertical motion of both gas and dust. Interestingly, we find a growth condition similar to the previous condition using total dust and gas surface densities regardless of the orders-of-magnitude difference in the gas and dust scale heights.

This paper is organized as follows. Section \ref{sec:basic_eqs} describes basic equations for dust and gas. In Section \ref{sec:linear_prob}, we describe setups of the eigenvalue problem, e.g., the background state and linearized equations. We present obtained growth rates and flow structures of dust and gas in Section \ref{sec:result}. In Section \ref{sec:discussion}, we discuss (1) the dependence of the boundary position of gas (Section \ref{subsec:zs_dependence}) and (2) a critical dust mass for secular GI to operate and dust mass budget that recent observations suggest (Section \ref{subsec:comp_obs}). We comment briefly on processes neglected in this work in Section \ref{subsec:visc_shear} and give a summary in Section \ref{sec:summary}.

\section{Basic equations}\label{sec:basic_eqs}

We describe equations for gas and dust and the assumed background state in detail in the following subsections. Here, we first briefly note our methodology and its motivation. 

We consider an axisymmetric dusty-gas disk around a central star whose mass is $M_{\ast}$. The main focus of this study is how much the large difference between the dust and gas vertical scales modifies the growth condition of secular GI. To focus on this issue, we utilize stratified shearing box coordinates $(x,y,z)$ \citep[][]{Goldreich1965b}. We assume the angular velocity of the shearing box to be the Keplerian angular velocity $\Omega$ at a radial distance from the central star $R$. 

We also make equations as simple as possible by neglecting the dust drifting motion at the unperturbed state \citep[][]{Weidenschilling1977,Nakagawa1986} and viscous momentum transport within the gas \citep[][]{Shakura1973}. The radial drift of dust grains does not change the growth condition according to the previous studies \citep[][]{Youdin2005a,Tominaga2020}. Our previous study showed that turbulent viscosity augments secular GI and also introduces another secular instability named two-component viscous GI \citep[TVGI,][]{Tominaga2019}. We note that turbulent diffusion of dust grains is an important process stabilizing secular GI, and thus we treat only the dust diffusion among the processes relevant to gas turbulence. Adopting this simplification, we can derive a sufficient condition for secular GI to operate even if we neglect those effects. This methodology is enough to test the conjecture about the stability of secular GI in the previous studies (see also Section \ref{subsec:visc_shear}). 

\subsection{Gas equations}
We use the following gas equations:
\begin{equation}
\frac{\partial \rhog}{\partial t} + \frac{\partial \rhog u_x}{\partial x} + \frac{\partial \rhog u_z}{\partial z} = 0,\label{eq:gas_eoc}
\end{equation}
\begin{align}
\frac{\partial u_x}{\partial t} + u_x\frac{\partial u_x}{\partial x} + u_z\frac{\partial u_x}{\partial z} =& -\frac{\cs^2}{\rhog}\frac{\partial \rhog}{\partial x} +3\Omega^2x +2\Omega u_y \notag\\
&-\frac{\partial\Phi}{\partial x} + \frac{\rhod}{\rhog}\frac{v_x-u_x}{\tstop},\label{eq:gas_eomx}
\end{align}
\begin{equation}
\frac{\partial u_y}{\partial t} + u_x\frac{\partial u_y}{\partial x} + u_z\frac{\partial u_y}{\partial z} = -2\Omega u_x + \frac{\rhod}{\rhog}\frac{v_y-u_y}{\tstop},\label{eq:gas_eomy}
\end{equation}
\begin{equation}
\frac{\partial u_z}{\partial t} + u_x\frac{\partial u_z}{\partial x} + u_z\frac{\partial u_z}{\partial z} = -\frac{\cs^2}{\rhog}\frac{\partial\rhog}{\partial z}-\Omega^2 z -\frac{\partial\Phi}{\partial z}+\frac{\rhod}{\rhog}\frac{v_z-u_z}{\tstop}, \label{eq:gas_eomz}
\end{equation}
where $\rho_{\gas}$ and $\rho_{\dst}$ are the gas and dust densities, $u_i,$ and $v_i$ for $i=x,\;y,\;z$ denote the gas and dust velocities in each direction, $\cs$ is the isothermal sound speed, $\tstop$ is the stopping time of dust. In this work, we assume $\tstop$ to be independent of $z$. Stricktly speaking, the stopping time depends on $\rho_{\gas}$, and $\tstop$ of a given dust size is smaller in the less dense region above the midplane. In this work, we focus on cases where the dust scale height is much smaller than the gas scale height because of weak turbulence. In these cases, the gas density is almost uniform within the dust layer, and the assumption of the uniform $\tstop$ is valid. The self-gravitational potential $\Phi$ is given by the Poisson equation:
\begin{equation}
\frac{\partial^2\Phi}{\partial x^2} + \frac{\partial^2\Phi}{\partial z^2} = 4\pi G(\rhog+\rhod). \label{eq:poisson}
\end{equation}

We assume that the gas is vertically isothermal for simplicity. The characteristic scale of gas is $H\equiv\cs/\Omega$, which gives the gas scale height in the absence of the self-gravity. Because of the axisymmetry, the azimuthal force includes only the Coriolis force and the aerodynamic friction. The second term on the right-hand side of Equation (\ref{eq:gas_eomz}) represents the stellar gravity under the assumption of $H/R\ll1$.

\subsection{Dust equations and diffusion modeling}
We describe equations for dust with turbulent diffusion in this subsection. One important process to discuss the growth condition of secular GI is turbulent diffusion of dust grains since the diffusion is the most efficient stabilizing process. This process is often modeled as a diffusion term in the continuity equation \citep[e.g.,][]{Dubrulle1995}:
\begin{equation}
\frac{\partial \rho_{\dst}}{\partial t}+\frac{\partial}{\partial x_j}\left(\rho_{\dst}v_j\right) = \frac{\partial}{\partial x_j}\left(D\rho_{\gas}\frac{\partial\rho_{\dst}/\rho_{\gas}}{\partial x_j}\right),
\end{equation}
where $D$ denotes the dust diffusivity, $j$ takes $x,\; y$, or $z$, and the repeated indices indicate the summation over them. Here, we assume the diffusion flux is proportional to the concentration gradient \citep[][]{Dubrulle1995} although there is another way using the density gradient to describe the diffusion \citep[][]{Cuzzi1993}. We found an insignificant difference in the properties of secular GI for parameters explored in this work. Although the above diffusion model has been widely adopted, we showed that simply adding the diffusion term in the continuity equation violates the total angular momentum conservation in the dust-gas system \citep{Tominaga2019}. The violation of the conservation law unphysically changes the mode properties of secular GI. For example, the above diffusion model leads to an overstable mode as secular GI which originates from the violation of the angular momentum conservation \citep[see Figure 2 in][]{Tominaga2019}. Therefore, we need to avoid this issue by modifying the dust equations.

So far, two possible ways have been proposed to recover the conservation law. One way is to add diffusion-related terms in the equation of motion for dust as well as in the continuity equation \citep[][]{Tominaga2019,Huang2022}. In our previous study \citep{Tominaga2019}, we showed that the total angular momentum is conserved if one includes the following diffusion velocity $v_{\mathrm{diff}}$ in the advection velocity of momentum of dust\footnote{In \citet{Tominaga2019}, we described the diffusion based on the density gradient following \citet{Cuzzi1993}. This difference does not affect the conclusion that the conservation law is recovered once one adds the diffusion velocity in the advection term.}
\begin{equation}
v_{\mathrm{diff},j}\equiv -D\frac{\rhog}{\rhod}\frac{\partial\rho_{\dst}/\rho_{\gas}}{\partial x_j}.
\end{equation}
\citet{Huang2022} extended our formalism and showed that further considering the time derivative of the diffusion velocity ensures the Galilean invariance. The essence of this modeling is to consider that both mass and momentum are carried by the diffusion flow. It is assumed that the momentum transfer between dust and gas occurs through the friction associated with the velocity difference of the mean flow, $v_i-u_i$.

The other way was proposed by \citet{Klahr2021}. They modeled the diffusion by using a pressure-like term in the equation of motion of dust. A set of their equations are the following:
\begin{equation}
\frac{\partial \rho_{\dst}}{\partial t}+\frac{\partial}{\partial x_j}\rho_{\dst}v_j =0,\label{eq:eoc_no_diffusion}
\end{equation}
\begin{equation}
\frac{\partial v_i}{\partial t} + v_j\frac{\partial v_i}{\partial x_j} = f_i -\frac{D\rho_{\gas}}{\tstop\rho_{\dst}}\frac{\partial}{\partial x_i}\left(\frac{\rho_{\dst}}{\rho_{\gas}}\right)-\frac{v_i-u_i}{\tstop},\label{eq:klahr_eom}
\end{equation}
where $f_i$ denotes external forces per unit mass. As shown in \citet{Klahr2021}, they derived the second term in Equation (\ref{eq:klahr_eom}) considering a specific case where $f_i$ is the gravity. The assumption they made is that the equations reproduce a steady state where (1) the velocity $v_i$ is zero and (2) dust sedimentation is balanced by the diffusion. We refer readers to Appendix B and C of \citet{Klahr2021} for more detailed description and discussion.

In this formalism, the velocity $v_i$ represents the sum of the mean-flow velocity and the diffusion velocity. This can be easily shown in the terminal velocity limit:
\begin{equation}
v_i = \tstop f_i -\frac{D\rho_{\gas}}{\rho_{\dst}}\frac{\partial\rho_{\dst}/\rho_{\gas}}{\partial x_i},
\end{equation}
where we omit the gas velocity just for simplicity and clarity. The first term on the right-hand side is the terminal velocity in the usual manner. The second term is the diffusion velocity, and we can readily see that this leads to the diffusion term in the continuity equation by substituting the terminal velocity in the continuity equation (Equation (\ref{eq:eoc_no_diffusion})), that is, 
\begin{equation}
\frac{\partial \rho_{\dst}}{\partial t}+\frac{\partial}{\partial x_j}\left(\rho_{\dst}\tstop f_j\right) = \frac{\partial}{\partial x_j}\left(D\rho_{\gas}\frac{\partial\rho_{\dst}/\rho_{\gas}}{\partial x_j}\right).
\end{equation}

It should be also noted that the advection velocity of their equation of motion (Equation (\ref{eq:klahr_eom})) implicitly includes the effect of the diffusion flow, which is the key ingredient to recover the angular momentum conservation as proposed in our previous study \citep{Tominaga2019}. Besides, the equation includes the time derivative of the diffusion velocity, which \citet{Huang2022} introduced to recover the Galilean invariance. In this way, the above two models proposed the similar equations. As in the former model, the key assumption for the momentum conservation is that the diffusion flow carries both mass and momentum of dust grains. In other words, this model also assumes that dust and gas exchange their momentums only through the friction due to the velocity difference of the mean flow. The dispersion relation of one-fluid secular GI derived based on \citet{Klahr2021} is consistent with the dispersion relation based on our modeling \citep[][]{Tominaga2020}\footnote{\citet{Tominaga2020} includes the radial drift of dust grains (see Section 4.1 therein). One has to compare the dispersion relation of \citet{Klahr2021} (Equation (B19) therein) and ours \citep[Equation (20) of][]{Tominaga2020} at the no-drift limit.}. Thus, both models can be used for linear analyses of secular GI. 

There are some merits of using the model of \citet{Klahr2021}. One merit is the simplicity of the equations as already noted in \citet{Klahr2021}. Another merit is more technical and is that $v_z$ becomes 0 in the settling-diffusion equilibrium. It is known that the non-zero vertical velocity of dust grains can cause dust settling instability at much small scales $\ll H$ \citep[e.g.,][]{Squire2018a,Zhuravlev2019,Krapp2020}, which is irrelevant to secular GI. We can eliminate the settling instability in effect from our analyses by using the model of \citet{Klahr2021}, and this allows us to present results more clearly.

We also note that one can treat streaming instability using the model of \citet{Klahr2021} if one assumes non-zero horizontal dust drift due to the global gas pressure gradient in the background state. As mentioned in the beginning of Section \ref{sec:basic_eqs}, we focus on secular GI by neglecting the background dust drift. Neglecting the background drift motion does not affect growth rates of secular GI (see Section \ref{subsec:visc_shear}).
 
In this way, we follow \citet{Klahr2021} in this work and utilize the following equations:
\begin{equation}
\frac{\partial \rhod}{\partial t} + \frac{\partial \rhod v_x}{\partial x} + \frac{\partial \rhod v_z}{\partial z} = 0,\label{eq:dust_eoc}
\end{equation}
\begin{align}
\frac{\partial v_x}{\partial t} + v_x\frac{\partial v_x}{\partial x} + v_z\frac{\partial v_x}{\partial z} =& -\frac{D_x\rhog}{\tstop\rhod}\frac{\partial }{\partial x}\left(\frac{\rhod}{\rhog}\right) +3\Omega^2x +2\Omega v_y\notag\\
& -\frac{\partial\Phi}{\partial x} -\frac{v_x-u_x}{\tstop},\label{eq:dust_eomx}
\end{align}
\begin{equation}
\frac{\partial v_y}{\partial t} + v_x\frac{\partial v_y}{\partial x} + v_z\frac{\partial v_y}{\partial z} = -2\Omega v_x - \frac{v_y-u_y}{\tstop},\label{eq:dust_eomy}
\end{equation}
\begin{align}
\frac{\partial v_z}{\partial t} + v_x\frac{\partial v_z}{\partial x} + v_z\frac{\partial v_z}{\partial z} =& -\frac{D_z\rhog}{\tstop\rhod}\frac{\partial}{\partial z}\left(\frac{\rhod}{\rhog}\right)-\Omega^2 z \notag\\
&-\frac{\partial\Phi}{\partial z}-\frac{v_z-u_z}{\tstop}, \label{eq:dust_eomz}
\end{align}
where $D_x$ and $D_z$ are the diffusivity in the radial and vertical directions. We adopt the same value for $D_x$ and $D_z$ in the fiducial case (the case 1 in Table \ref{tab:params}). We can discuss possible impact of the vertical thickness on secular GI by changing the ratio $D_z/D_x$. 

In this work, we assume the dust diffusivity to be constant and independent from $\rhod/\rhog$ for simplicity. However, the diffusivity can depend on $\rhod/\rhog$ especially when the dust density is much larger than $\rhog$ so that feedback from dust to gas turbulence is important \citep[e.g.,][]{Takeuchi2012,Schreiber2018,Xu2022,Gerbig2023}. The previous studies found that the diffusivity decreases as $\rhod/\rhog$ increases. \citet{Michikoshi2012} showed that density-dependent diffusion can enhance the growth of instabilities. Our analyses thus provide sufficient conditions for secular GI.

\section{Formulation of eigenvalue problem}\label{sec:linear_prob}

\subsection{Background state and numerical domain}
We denote unperturbed background state values with a subscript ``0". We assume a radially uniform density profile and also assume $u_{x,0}=v_{x,0}=0$. The background state is then given by the following equations including one dimensional ordinary differential equations (ODEs):
\begin{equation}
u_{z,0}=v_{z,0}=0,
\end{equation}
\begin{equation}
u_{y,0}=v_{y,0}=-\frac{3}{2}\Omega x,
\end{equation}
\begin{equation}
-\frac{\cs^2}{\rhogup}\frac{d\rhogup}{d z}-\Omega^2 z -\frac{d\Phi_0}{d z}=0, \label{eq:gas_eomz_unp}
\end{equation}
\begin{equation}
-\frac{D_z\rhogup}{\tstop\rhodup}\frac{d}{d z}\left(\frac{\rhodup}{\rhogup}\right)-\Omega^2 z -\frac{d\Phi_0}{d z}=0, \label{eq:dust_eomz_unp}
\end{equation}
\begin{equation}
\frac{d\Phi_0}{dz} = g_{z,0}, \label{eq:gz_unp}
\end{equation}
\begin{equation}
\frac{dg_{z,0}}{d z} = 4\pi G(\rhogup+\rhodup). \label{eq:poisson_unp}
\end{equation}

In the absence of the self-gravity, the unperturbed density structures are given by the following:
\begin{equation}
\rhogup(z)=\rhogup(0)\exp\left(-\frac{z^2}{2H^2}\right)
\end{equation}
\begin{equation}
\rhodup(z)=\rhodup(0)\exp\left(-\frac{z^2}{2\hd'^2}\right),
\end{equation}
\begin{equation}
\epsilon(z)=\epsilon(0)\exp\left(-\frac{z^2}{2\hd^2}\right),
\end{equation}
where
\begin{equation}
\epsilon(z)\equiv\rhodup/\rhogup,
\end{equation}
\begin{equation}
\hd'^{-2} = H^{-2}+\hd^{-2},
\end{equation}
\begin{equation}
\hd\equiv\sqrt{\frac{D_z}{\tstop\Omega^2}}.
\end{equation}
The vertical scale $\hd'$ is almost equal to $\hd$ for sufficiently weak turbulence. In the presence of the self-gravity, the background state is characterized by four dimensionless parameters: (1) the midplane dust-to-gas ratio, $\epsilon(0)$, (2) the dimensionless vertical diffusivity, $\tilde{D}_z\equiv D_{z}/\cs H$, (3) the dimensionless stopping time, $\taus\equiv\tstop\Omega$, and (4) the 3D analogue of Toomre's parameter for gas \citep[e.g.,][]{Goldreich1965a,Mamatsashvili2010}:
\begin{equation}
\qthree\equiv\frac{\Omega^2}{4\pi G\rhogup(0)}.
\end{equation}
The parameter $\qthree$ is equal to $Q\sqrt{2\pi}/4\simeq 0.6Q$, where $Q\equiv\cs\Omega/\pi G \Sigma_{\gas}$ if the vertical gas structure is exactly given by the Gaussian profile with the scale height of $H$, i.e., $\sqrt{2\pi}H\rhogup(0)=\Sigma_{\gas}$. A more robust relation between $Q$ and $\qthree$ is given in \citet{Mamatsashvili2010} and \citet{Lin2014}. We consider $\qthree\geq1.5$ for which a gas disk is stable to GI \citep[e.g., see][]{Goldreich1965a,Sekiya1983,Mamatsashvili2010}. Sets of the adopted parameter values are summarized in Table \ref{tab:params}. 

Using the fourth-order Runge-Kutta integrator, we numerically solve Equations (\ref{eq:gas_eomz_unp})-(\ref{eq:poisson_unp}) to obtain the unperturbed densities and the gravitational potential. We consider the symmetric profile with respect to the midplane and solve the above ODEs from $z=0$ to $z=\zs >0$. The inner boundary conditions at $z=0$ for $g_{z,0}$ to obtain the symmetric profile is $g_{z,0}(0)=0$. Because of the assumption of the isothermal gas, there is no exact outer boundary where the gas density becomes zero. We thus set $\zs=2H$ and truncate the profiles at the boundary (see Section \ref{subsec:zs_dependence}). To conduct numerical integration, we use 513 grid points including $z=0$, and 512 of them are logarithmically spaced from $z=10^{-2}\hd$ to $z=\zs$. We tested the convergence of the growth rates and found that error is less than 0.1 percent when we adopt more than 100 grids.\footnote{We conducted the convergence test by changing the number of grid points for $10^{-2}\hd\leq z\leq\zs$. We measured the error with respect to the growth rate obtained with 1024 grids.} 

\begin{figure}[tp]%[htp] or [H]
	\begin{center}
	\hspace{100pt}\raisebox{20pt}{
	\includegraphics[width=0.8\columnwidth]{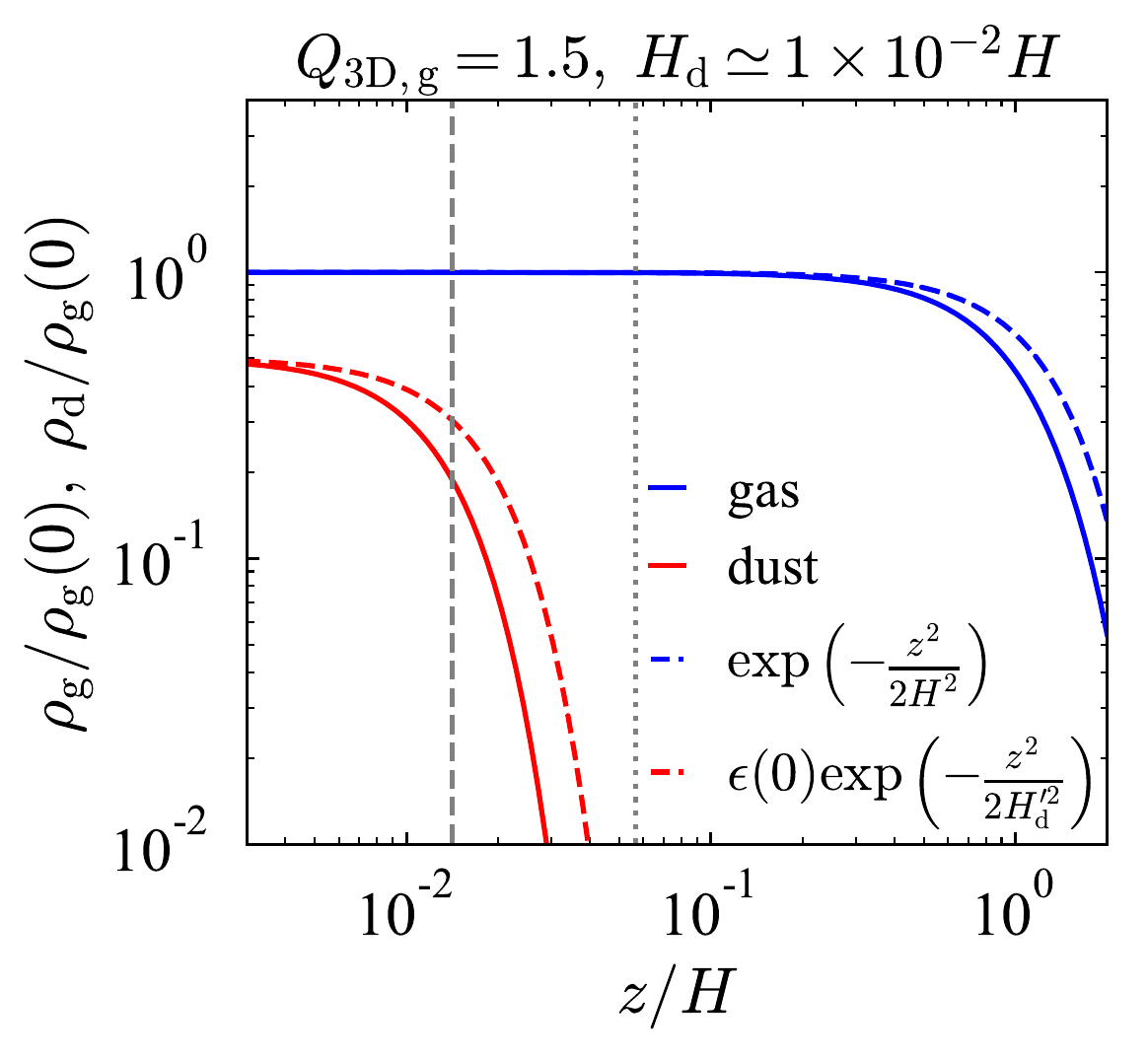} %2 column
%	\hspace{0pt}\raisebox{20pt}{
%	\includegraphics[width=0.5\columnwidth]{./disp_Q3d1.5_alpr1e-4_alpz1e-4_taus05_dgmid05_Hs2Hg_unp_dens_prof.pdf} %1 column
	}
	\end{center}
	\vspace{-30pt}
\caption{The unperturbed dust (red solid line) and gas densities (blue solid line) for the case 1  in Table \ref{tab:params}. Both densities are normalized by the midplane gas density. The dashed lines represent the solutions of the gas and dust densities in the absence of the self-gravity. The vertical scale of the latter profile (red dashed line) is given by $\hd '^{-2}=H^{-2}+\hd^{-2}\simeq\hd^{-2}$. The vertical dashed and dotted lines mark $z=\hd$ and $z=4\hd$, respectively.} 
\label{fig:unp_strat_example}
\end{figure}

Figure \ref{fig:unp_strat_example} shows the unperturbed density profiles for the case 1 in Table \ref{tab:params}. The gas density is almost uniform within the dust layer. Because of the self-gravity, the dust and gas densities deviate from the Gaussian profile expected in the absence of the self-gravity. We define the dust and gas surface densities as follows
\begin{equation}
\Sigma_{\gas}\equiv 2\int_0^{\zs}\rhogup dz,
\end{equation}
\begin{equation}
\Sigma_{\dst}\equiv 2\int_0^{\zs}\rhodup dz,
\end{equation}
and also define the dust-to-gas surface density ratio:
\begin{equation}
\varepsilon\equiv \frac{\Sigma_{\dst}}{\Sigma_{\gas}}.
\end{equation}
Adopting these surface densities, $\tstop$, $D_r$, and $\hd$ for the previous razor-thin-disk analysis, we can calculate the growth rate of 1D secular GI. We use those 1D growth rates to compare the growth rates obtained in the present study. 

In this study, we consider vertically thin dust disks whose vertical scale height is much smaller than $H$. In this case, the unperturbed dust density is negligibly small in the upper layer, and such dust grains much above the dust layer are not important from the dynamical point of view. However, this too-small dust density can cause a numerical issue that the dust velocity perturbations diverge in the upper region when we numerically solve the eigenvalue problem described in the next section. We thus set another outer boundary at $z=\zsd$ and enforce the dust density to be zero at $z>\zsd$. We adopt $\zsd=4\hd$ in this study. We confirmed that truncating dust profiles at $z=4\hd$ makes insignificant differences in growth rates compared to results in a case where we take the small dust density in the upper layer into account adopting $\zsd=\zs$ with sufficiently high numerical resolutions. 

% ============== Table ================================

\begin{deluxetable*}{ccccccccc}[ht]
\tablecaption{Sets of parameters and growth rates} \label{tab:params}
\tablehead{
\colhead{Case} & 
\colhead{$\epsilon(0)$\tablenotemark{a}} &
\colhead{$\tilde{D_r}$} &
\colhead{$\tilde{D_z}$} & 
\colhead{$\taus$} & 
\colhead{$\qthree$} &
\colhead{$kH$\tablenotemark{b}} &
\colhead{$n/\Omega$} & 
\colhead{$n_{\mathrm{1d}}/n$\tablenotemark{c}} 
}
\startdata
1 &
0.5 &
1.0$\times 10^{-4}$ &
1.0$\times 10^{-4}$ &
0.5 &
1.5 &
8.7 &
$3.05\times10^{-3}$ &
1.47 \\ 
2 &
0.5 &
1.0$\times 10^{-4}$ &
1.0$\times 10^{-4}$ &
0.1 &
1.5 &
3.9 &
$2.77\times10^{-4}$ &
1.04 \\ 
3 &
0.5 &
1.0$\times 10^{-4}$ &
3.0$\times 10^{-4}$ &
0.2 &
1.5 &
7.3 &
$2.79\times10^{-3}$ &
1.49 \\ 
4 &
0.3 &
3.0$\times 10^{-4}$ &
1.0$\times 10^{-3}$ &
0.2 &
1.5 &
3.0 &
$1.35\times10^{-4}$ &
-1.95 \\ 
5 &
0.3 &
1.0$\times 10^{-4}$ &
1.0$\times 10^{-3}$ &
0.5 &
2.0 &
8.5 &
$4.19\times10^{-3}$ &
1.27 \\ 
6 &
0.5 &
1.0$\times 10^{-4}$ &
1.0$\times 10^{-4}$ &
0.5 &
2.0 &
7.1 &
$1.42\times10^{-3}$ &
1.44 \\ 
7 &
2.0 &
3.0$\times 10^{-5}$ &
3.0$\times 10^{-5}$ &
0.05 &
3.0 &
8.8 &
$3.14\times10^{-4}$ &
3.32 \\ 
\enddata
\tablenotetext{a}{The midplane dust-to-gas ratio: $\epsilon(0)=\rhodup(0)/\rhogup(0)$}
\tablenotetext{b}{We pick the wavenumber close to the most unstable wavenumber.}
\tablenotetext{c}{The ratio of the 1D growth rate $n_{\mathrm{1D}}$ obtained in the thin-disk framework \citep{Tominaga2019} and the growth rate $n$ in this work}
\end{deluxetable*}
% ============== Table ================================

\subsection{Linearized equations}\label{subsec:linear_eqs}

We assume sinusoidal perturbations in the radial direction upon the background state. Perturbation of a physical value $A$ has the form of $\delta A(z)\exp(ikx+nt)$ where $n$ is complex growth rate and $k$ is wavenumber. We then obtain the following linearized equations:
\begin{equation}
n\delta\rhog+ik\rhogup\delta u_x + \rhogup \frac{d\delta u_z}{d z} + \delta u_z\frac{d\rhogup}{d z}=0,\label{eq:gas_eoc_linear}
\end{equation}
\begin{equation}
n\delta u_x = -ik\cs^2\frac{\delta\rhog}{\rhogup} + 2\Omega\delta u_y - ik\delta\Phi + \frac{\rhodup}{\rhogup}\frac{\delta v_x-\delta u_x}{\tstop},\label{eq:gas_eomx_linear}
\end{equation}
\begin{equation}
n\delta u_y = -\frac{\Omega}{2}\delta u_x + \frac{\rhodup}{\rhogup}\frac{\delta v_y-\delta u_y}{\tstop},\label{eq:gas_eomy_linear}
\end{equation}
\begin{equation}
n\delta u_z = -\frac{\cs^2}{\rhogup}\frac{d \delta\rhog}{d z}  +\frac{\delta\rhog}{\rhogup}\frac{\cs^2}{\rhogup}\frac{d\rhogup}{dz}  - \frac{d\delta\Phi}{d z}  + \frac{\rhodup}{\rhogup}\frac{\delta v_z-\delta u_z}{\tstop}, \label{eq:gas_eomz_linear}
\end{equation}
\begin{equation}
n\delta\rhod+ik\rhodup\delta v_x +   \frac{d\rhodup\delta v_z}{d z} =0,\label{eq:dust_eoc_linear}
\end{equation}
\begin{equation}
n\delta v_x = -ik\frac{D_x}{\tstop}\frac{\delta\rhod}{\rhodup}+ik\frac{D_x}{\tstop}\frac{\delta\rhog}{\rhogup} + 2\Omega\delta v_y - ik\delta\Phi - \frac{\delta v_x-\delta u_x}{\tstop},\label{eq:dust_eomx_linear}
\end{equation}
\begin{equation}
n\delta v_y = -\frac{\Omega}{2}\delta v_x - \frac{\delta v_y-\delta u_y}{\tstop},\label{eq:dust_eomy_linear}
\end{equation}
\begin{align}
n\delta v_z = - \frac{D_z}{\tstop\rhodup}\frac{d \delta\rhod}{dz}  +\frac{\delta\rhod}{\rhodup}\frac{D_z}{\tstop\rhodup}\frac{d\rhodup}{dz}\notag\\
+ \frac{D_z}{\tstop\rhogup}\frac{d \delta\rhog}{dz}  -\frac{\delta\rhog}{\rhogup}\frac{D_z}{\tstop\rhogup}\frac{d\rhogup}{dz}\notag\\
- \frac{d\delta\Phi}{d z}  -\frac{\delta v_z-\delta u_z}{\tstop}, \label{eq:dust_eomz_linear}
\end{align}
\begin{equation}
-k^2\delta\Phi + \frac{d^2\delta\Phi}{d z^2} = 4\pi G(\delta\rhog+\delta\rhod). \label{eq:Poisson_linear}
\end{equation}
Rearranging these equations with $\delta M_{\dst,x}\equiv\rhodup\delta v_x$ and $\delta M_{\dst,z}\equiv \rhodup\delta v_z$, we obtain the following six ODEs:
\begin{align}
\frac{d\delta u_z}{dz} = -n\frac{\delta\rhog}{\rhogup}-ik\delta u_x - \frac{d\ln\rhogup}{dz}\delta u_z,
\end{align}
\begin{align}
\frac{d\delta\rhog}{dz} = \frac{\rhogup}{\cs^2}\left(-n\delta u_z -\delta g_z + \frac{\delta M_{\dst,z}-\rhodup\delta u_z}{\rhogup\tstop}\right)\notag\\+ \frac{d\ln\rhogup}{dz}\delta \rhog,
\end{align}
\begin{align}
\frac{d\delta M_{\dst,z}}{dz} = - n\delta \rhod - ik\delta M_{\dst,x},
\end{align}
\begin{align}
\frac{d\delta\rhod}{dz} = &\frac{\tstop}{D_z}\left[-n\delta M_{\dst,z}-\rhodup\delta g_z-\frac{\delta M_{\dst,z}-\rhodup\delta u_z}{\tstop}\right] \notag\\
&+\frac{d\ln\rhodup}{dz}\delta\rhod+ \frac{\rhodup}{\rhogup}\frac{d\delta\rhog}{dz}-\frac{\rhodup}{\rhogup}\frac{d\ln\rhogup}{dz}\delta\rhog,
\end{align}
\begin{align}
\frac{d\delta\Phi}{dz}=\delta g_z,
\end{align}
\begin{align}
\frac{d\delta g_z}{dz} = 4\pi G(\delta\rhog+\delta\rhod)+k^2\delta\Phi,
\end{align}
\begin{align}
i\delta u_x = W_2(n)k\delta\Phi + \frac{\rhodup W_1(n)+\rhogup W_2(n)}{\rhogup(\rhogup+\rhodup)}\cs^2 k\delta\rhog\notag\\
-\frac{W_1(n)-W_2(n)}{\rhogup+\rhodup}\frac{D_x}{\tstop}k\left(\delta\rhod - \frac{\rhodup}{\rhogup}\delta\rhog\right), 
\end{align}
\begin{align}
i\delta M_{\dst,x}=\rhodup W_2(n) k\delta\Phi - \frac{\rhodup\left(W_1(n)-W_2(n)\right)}{\rhogup+\rhodup}\cs^2 k \delta\rhog \notag \\
+\frac{\rhogup W_1(n) + \rhodup W_2(n)}{\rhogup+\rhodup}\frac{D_x}{\tstop}k\left(\delta\rhod-\frac{\rhodup}{\rhogup}\delta\rhog\right),
\end{align}
where $W_1(n)$ and $W_2(n)$ are functions of $n$:
\begin{equation}
W_1(n)\equiv  \left(n+\frac{1+\rhodup/\rhogup}{\tstop}\right)\left[\left(n+\frac{1+\rhodup/\rhogup}{\tstop}\right)^2+\Omega^2\right]^{-1},
\end{equation}
\begin{equation}
W_2(n)\equiv n\left(n^2+\Omega^2\right)^{-1}.
\end{equation}
We numerically integrate the above ODEs from $z=0$ toward $z=\zs$ using the fourth-order Runge-Kutta integrator and seek an eigenvalue $n$ utilizing the shooting method. The grid points are the same as the one we use to numerically derive the unperturbed state in the previous subsection. We normalize the physical values using $\rhogup(0)$, $\cs$, and $\Omega$ when we conduct numerical integration. Those values are used to show results in the following sections. The parameters in this eigenvalue problem are dimensionless radial diffusivity, $\tilde{D}_r\equiv D_r/\cs H$, wavenumber, $kH$, and the parameters used to derive the unperturbed state (e.g., $\taus$, see Table \ref{tab:params}).

\subsection{Boundary conditions}
We can obtain the eigenvalue of the above ODEs with six appropriate boundary conditions. We adopt three inner boundary conditions as follows
\begin{equation}
\delta u_z(0)=0,
\end{equation}
\begin{equation}
\delta v_z(0)=0,
\end{equation}
\begin{equation}
\delta g_z(0)=0,
\end{equation}
with which we can obtain even modes where the density profile is symmetric with respect to the midplane. We set three outer boundary conditions. One is the free surface condition for gas under the constant external pressure \citep[see][]{Goldreich1965a}:
\begin{equation}
n\delta\rhog(\zs) -\rhogup (\zs)(\Omega^2 \zs+g_{z,0}(\zs))\delta u_z(\zs)=0. \label{eq:freeBC_gas}
\end{equation}
We adopt this condition since we assume the surface ($z=\zs$) of the isothermal gas disk that would otherwise vertically extends infinitely. One can derive the above condition by assuming the Lagrangian perturbation of gas pressure, $\Delta(\cs^2\rhog)$, to be zero at the boundary $z=\zs$ \citep[e.g., see][]{Lin2015}:
\begin{equation}
\Delta(\cs^2\rhog)\equiv \cs^2\delta\rhog(\zs) + \frac{\delta u_z(\zs)}{n}\frac{d(\cs^2\rhogup)}{dz}\Bigg|_{\zs}=0.
\end{equation}
We note that smaller-$\zs$ cases correspond to high-external-pressure cases because of the negative vertical gradient of the gas pressure. 

The second condition is that the dust vertical velocity is zero at the outer boundary of the dust disk ($z=\zsd$):
\begin{equation}
\delta v_z(\zsd)=0.
\end{equation}
The reason to adopt this condition is as follows. As mentioned in the previous subsection, the dust density is substantially small in the upper region, which can lead to divergence of $\delta v_z$ in numerical calculations regardless of the finite value of $\delta M_{\dst,z}=\rhodup\delta v_z$. To avoid this, we assume $\delta v_z$ to be zero at the outer boundary. We have confirmed that the choice of this boundary condition or the location to impose this boundary condition does not significantly change the present results.

The last condition is for the self-gravity. The self-gravity and its potential must be continuous at the perturbed disk surface \citep{Nagai1998}. This gives the condition on $\delta\Phi$ and $\delta g_z=d\delta\Phi/dz$ as follows
\begin{equation}
-k\delta\Phi(\zs) - \delta g_z(\zs)=4\pi G\rhogup(\zs)\frac{\delta u_z(\zs)}{n},\label{eq:SG_obc}
\end{equation}
which is identical to a condition introduced in \citet{Goldreich1965a}. They derived the condition using (1) the continuity of potential, (2) the Gauss's flux theorem with an ansatz on gas surface density (see their Equation (51)), and (3) the solution of the Laplace equation outside the disk \citep[see also][]{Lubow1993,Mamatsashvili2010}. We note that the right-hand side of Equation (\ref{eq:SG_obc}) includes only the gas properties since we assume that the dust is absent at $z>\zsd$.

\begin{figure}[t!]%[htp] or [H]
	\begin{center}
	\hspace{100pt}\raisebox{20pt}{
	\includegraphics[width=\columnwidth]{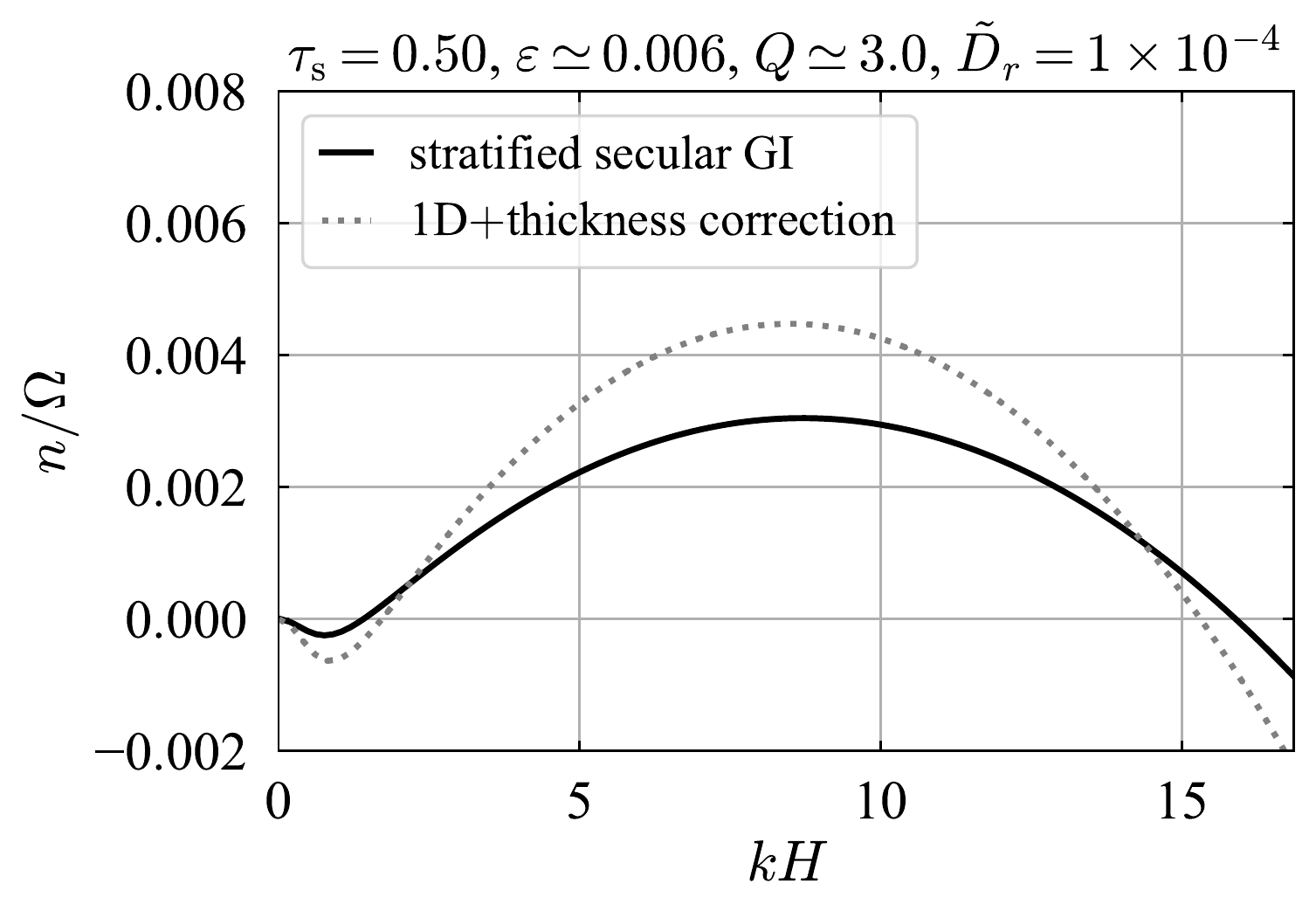} %2 column
%	\hspace{0pt}\raisebox{20pt}{
%	\includegraphics[width=0.5\columnwidth]{./disp_Q3d1.5_alpr1e-4_alpz1e-4_taus05_dgmid05_Hs2Hg_kH-ngrow_w_1D_disp.pdf} %1 column
	}
	\end{center}
	\vspace{-30pt}
\caption{Dispersion relation of stratified secular GI (black line) for the case 1. The gray dashed line is the dispersion relation obtained in the previous 1D analysis with the thickness correction term in the linearized Poisson equation \citep[][]{Vandervoort1970,Shu1984}. We refer readers to \citet{Tominaga2019} for the detail.} 
\label{fig:growth_rate_fiducial}
\end{figure}

\section{Results}\label{sec:result}
We first overview the obtained growth rates, dust- and gas-flow structures in Section \ref{subsec:ngrow_flow} and explain the driving force of the flow in Section \ref{subsec:force}. In Section \ref{subsec:anistropic_D}, we show another type of flow found in the anisotropic diffusion cases. Section \ref{subsec:cond} addresses the growth condition of secular GI in a vertically stratified disk. We also find another mode whose growth rate depends on the spatial resolution, but we neglect it in this study since its growth rate gradually decreases as we increase the resolution.

\begin{figure*}[t!]%[htp] or [H]
	\begin{center}
	\hspace{0pt}\raisebox{20pt}{
	\includegraphics[width=1.5\columnwidth]{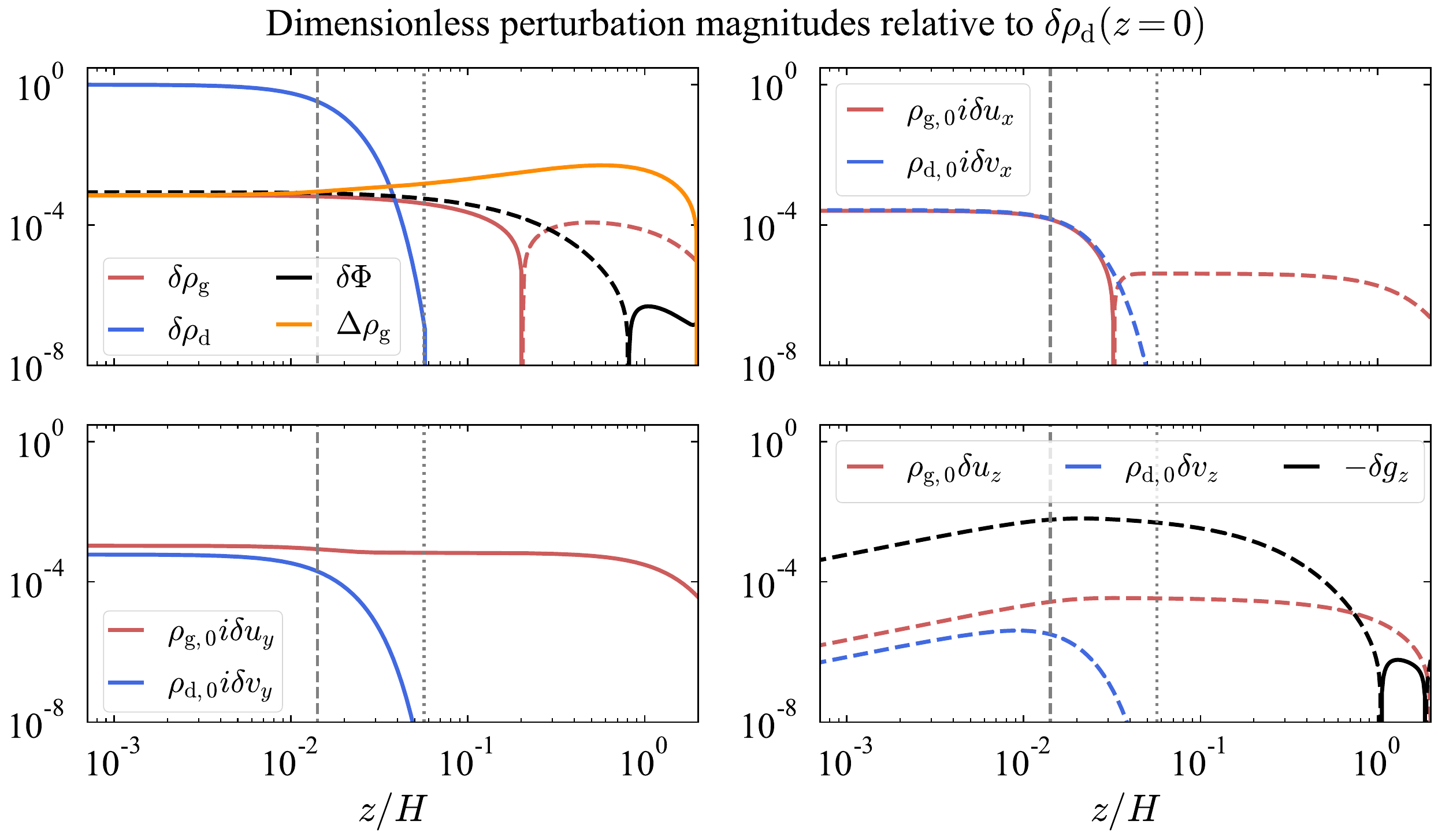} %2 column
%	\hspace{0pt}\raisebox{20pt}{
%	\includegraphics[width=0.9\columnwidth]{./disp_Q3d1.5_alpr1e-4_alpz1e-4_taus05_dgmid05_Hs2Hg_eigen_func_vertical_log_w_Lag.pdf} %1 column
	}
	\end{center}
	\vspace{-30pt}
\caption{Vertical profiles of the perturbed physical values in the case 1: $\delta\rhog$, $\delta\rhod$, $\delta\Phi$, and $\Delta\rhog$ in the top left panel, the radial (azimuthal) mass fluxes in the top right (bottom left) panels, and the vertical mass flux and $-\delta g_z$ in the bottom right panel. The plotted values are normalized by the midplane dust density perturbation $\delta\rhod(0)$, the isothermal sound speed $\cs$, and the Keplerian angular velocity $\Omega$. In each panel, the solid lines represent positive values while the dashed lines represent negative values. For example, $\delta\rhog$ changes its sign across $z\simeq0.2H$ (see the top left panel). It should be noted that $\Delta\rhog$ is positive throughout the domain. The vertical gray dashed and dotted lines mark $z=\hd$ and $z=\zsd=4\hd$, respectively.} 
\label{fig:eigen_func_fiducial}
\end{figure*}

\subsection{Growth rate and flow structures}\label{subsec:ngrow_flow}
Figure \ref{fig:growth_rate_fiducial} shows the obtained growth rate as a function of $kH$ for the case 1. We also show the dispersion relation obtained in the thin-disk framework given in \citet{Tominaga2019}, where we modify the linearized Poisson equation to mimic the effect of the finite disk thickness \citep[][]{Vandervoort1970,Shu1984}. We find that the most unstable wavenumber is similar to the previous 1D analysis. The difference in the maximum growth rates ($kH\simeq9$) is only a factor of $\simeq1.5$. We show growth rates for selected modes and parameters in Table \ref{tab:params}. The difference in the growth rate is within a factor of $\sim3$ for those parameter sets. We also find a case where stratified secular GI operates while the previous 1D secular GI is stable (the case 4), which is described in Section \ref{subsec:anistropic_D}

\begin{figure*}[t!]%[htp] or [H]
	\begin{center}
	\hspace{0pt}\raisebox{20pt}{
	\includegraphics[width=2.0\columnwidth]{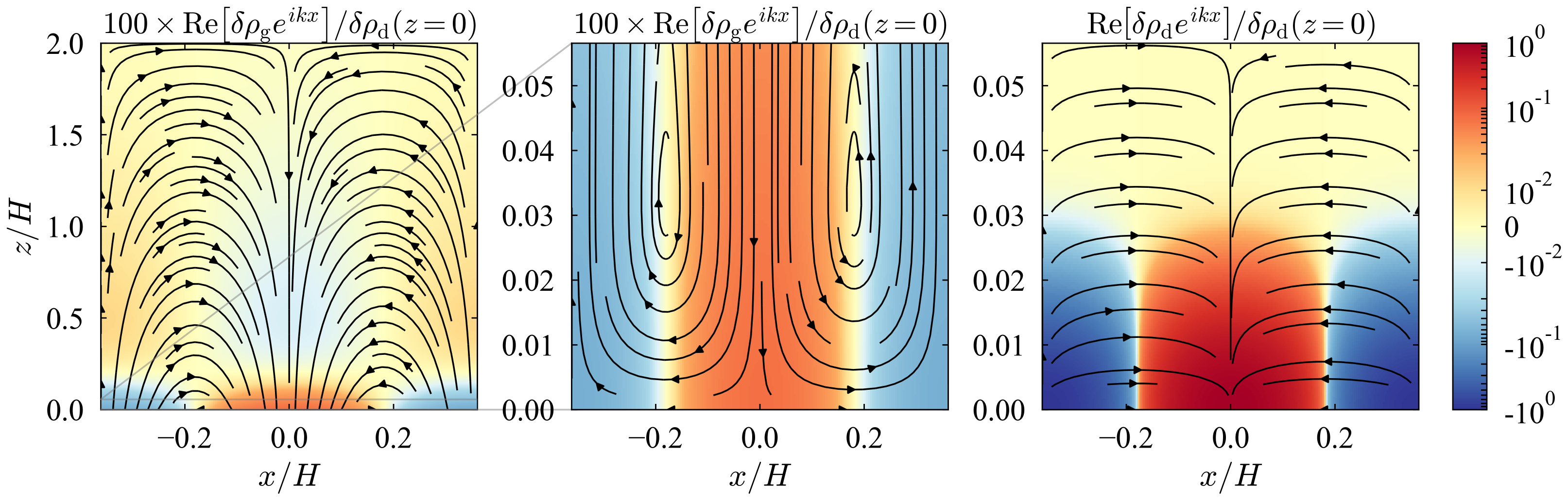} %2 column
%	\hspace{0pt}\raisebox{20pt}{
%	\includegraphics[width=0.9\columnwidth]{./disp_Q3d1.5_alpr1e-4_alpz1e-4_taus05_dgmid05_Hs2Hg_gas_and_dust_density_perturbation_three_panels.pdf} %1 column
	}
	\end{center}
	\vspace{-30pt}
\caption{Streamlines of gas (the left and middle panels) and dust (the right panel) in the case 1. The left panel shows the whole flow structure on the vertical scale of $2H$. The middle panel is the close-up view of the left panel and focuses on the dust layer ($0\leq z\leq \zsd$) as in the right panel. The color shows $100\mathrm{Re}\left[\delta\rhog e^{ikx}\right]/\delta\rhod(0)$ and $\mathrm{Re}\left[\delta\rhod e^{ikx}\right]/\delta\rhod(0)$ in the left and middle panels and in the right panel, respectively.} 
\label{fig:flow_struct_fiducial}
\end{figure*}

Figure \ref{fig:eigen_func_fiducial} shows the obtained eigenfunctions in the case 1. We use solid (dashed) lines to represent positive (negative) values of each physical variable. We also show 2D maps of the gas and dust density perturbations and their streamlines in Figure \ref{fig:flow_struct_fiducial}. The dust density perturbation is orders of magnitude larger than the gas density perturbation, which is similar to the previous 1D secular GI \citep[e.g., see][]{Tominaga2018,Tominaga2020}. Thus, the perturbed self-gravity is mostly due to the dust density perturbation, and the self-gravity of the perturbed gas is insignificant. The density and velocity perturbations of dust are confined within the dust layer. Interestingly, the gas density and velocity perturbations extend up to the outer gas boundary. 

There is a significant difference between the flow structures of gas and dust (see Figure \ref{fig:flow_struct_fiducial}). The dust motion is almost radial and toward the local maximum of $\delta\rhod$. The gas, however, shows meridional motion. The radial gas motion around the midplane is opposite to the dust motion, i.e., in the direction away from the dust density maximum. The vertical gas motion is in a direction to amplify the gas density perturbation at the dust density maximum. The gas supply due to the vertical motion at $x=0$ slightly dominates over the gas depletion due to the radial motion, resulting in the positive gas density perturbation that is in phase with $\delta\rhod$. We thus attribute the very small amplitude of $\delta\rhog/\delta\rhod(z=0)$ to the meridional circulation of gas. We note that the Lagrangian perturbation of the gas density $\Delta\rhog$ given below is positive throughout the vertical domain while the Eulerian perturbation $\delta\rhog$ is negative in the upper region:
\begin{equation}
\Delta\rhog\equiv\delta\rhog + \frac{d\rhogup}{dz}\frac{\delta u_z}{n}.
\end{equation}
Thus, gas fluid parcels become denser while they move toward the midplane, which makes the center of mass lower.

The much small amplitude of $\delta\rhog$ means that the gas motion is nearly incompressible during the linear growth of secular GI, which is reasonable. The timescale of secular GI is much longer than the Keplrian period (see Figure \ref{fig:growth_rate_fiducial}). On the other hand, the sound crossing time for the vertical scale of $\sim H$ is on the order of the Keplerian period and even shorter within the dust layer by a factor of $\sim \hd/H$. Thus, the sound wave can travel many times over the full vertical extent of the gas disk within one growth timescale of secular GI. In such a case, the gas behaves as if it is incompressible \citep[e.g., see][]{Landau1987}.

\begin{figure}[t!]%[htp] or [H]
	\begin{center}
	\hspace{100pt}\raisebox{20pt}{
	\includegraphics[width=0.8\columnwidth]{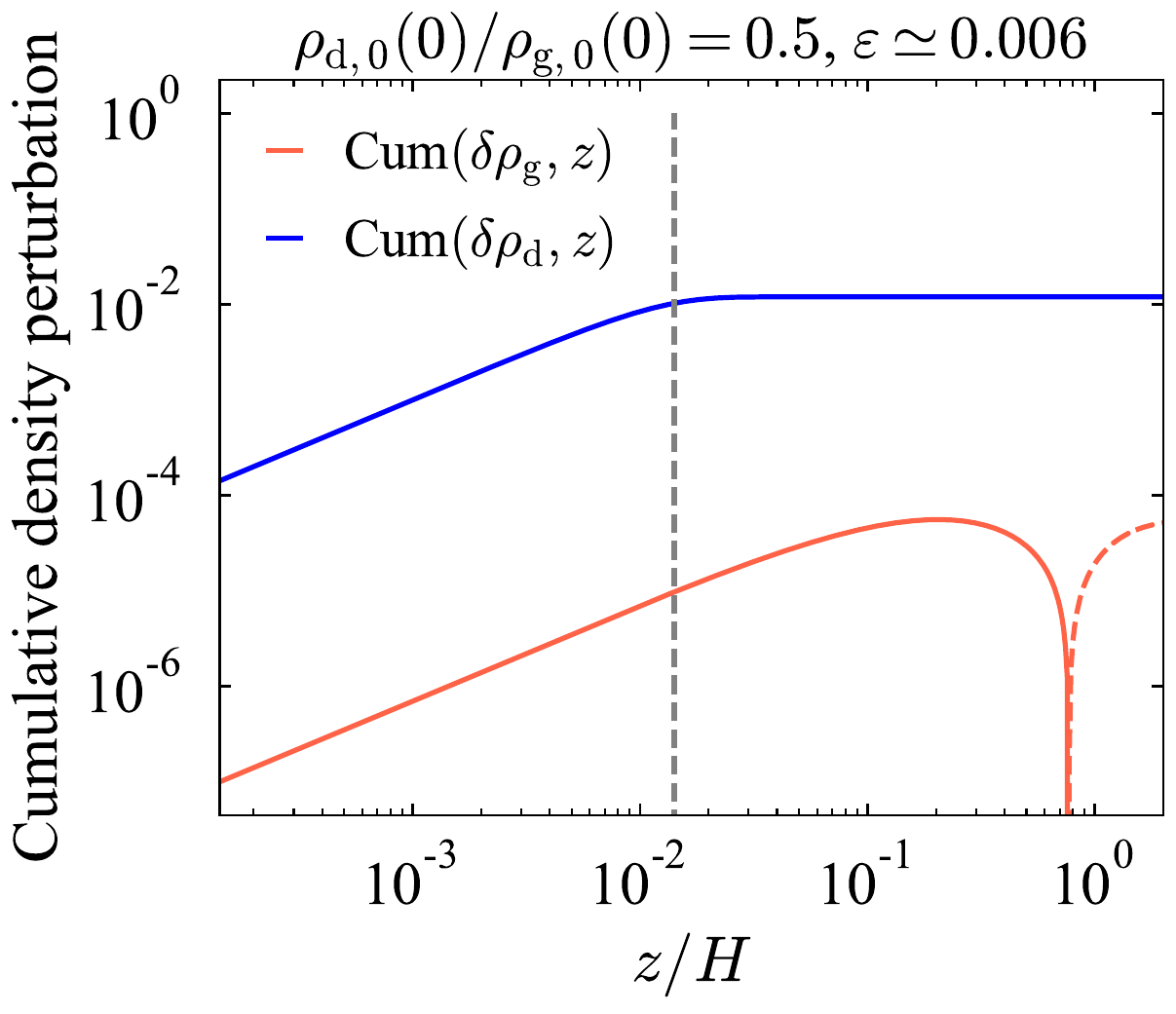} %2 column
%	\hspace{0pt}\raisebox{20pt}{
%	\includegraphics[width=0.5\columnwidth]{./disp_Q3d1.5_alpr1e-4_alpz1e-4_taus05_dgmid05_Hs2Hg_cumulative_density_perturbations.pdf} %1 column
	}
	\end{center}
	\vspace{-30pt}
\caption{Cumulative density perturbations in the case 1. Both density perturbations are normalized by $\delta\rhod(0)$. As in Figure \ref{fig:eigen_func_fiducial}, the solid line represents the region of a positive value while the dashed line represents the region of a negative value for each physical value. The vertical gray dashed line marks $z=\hd$. We find that the surface density perturbations of dust and gas are anti-phase.} 
\label{fig:cum_dens}
\end{figure}

%It should be noted that the sign of the gas perturbation changes because of the meridional motion. 
As mentioned above, there is a node in $\delta\rhog(z)$ at $z\simeq0.2H$, and the gas density perturbation $\delta\rhog$ is negative well above the dust layer (see Figure \ref{fig:eigen_func_fiducial}). We find that the region of negative $\delta\rhog$ is large and this leads to anti-phase dust and gas surface density perturbations. Figure \ref{fig:cum_dens} shows the cumulative density perturbations of dust and gas in the unit of $\delta\rho_{\dst}(0) H$. We define the cumulative value of a perturbation $\delta A$ as 
\begin{equation}
\mathrm{Cum}(\delta A,z)\equiv\int_0^{z}\delta Adz
\end{equation}
The surface density perturbations are given by
\begin{equation}
\delta\Sigma_{\gas} \equiv 2\mathrm{Cum}(\delta\rhog,\zs),
\end{equation}
\begin{equation}
\delta\Sigma_{\dst} \equiv 2\mathrm{Cum}(\delta\rhod,\zs)=2\mathrm{Cum}(\delta\rhod,\zsd).
\end{equation}
As in Figure \ref{fig:eigen_func_fiducial}, we use the dashed line to highlight a region where a perturbed value is negative. Figure \ref{fig:cum_dens} shows that $\delta\Sigma_{\dst}$ and $\delta\Sigma_{\gas}$ are in anti-phase, meaning that secular GI in a stratified disk creates a dust ring at a gap of the gas surface density. This surface density structure does not appear in the previous 1D analysis. Therefore, taking the vertical structure and motion into consideration is important to discuss the expected substructures due to secular GI.

\subsection{Driving force}\label{subsec:force}
\begin{figure}[t!]%[htp] or [H]
	\begin{center}
	\hspace{100pt}\raisebox{20pt}{	
	\includegraphics[width=0.9\columnwidth]{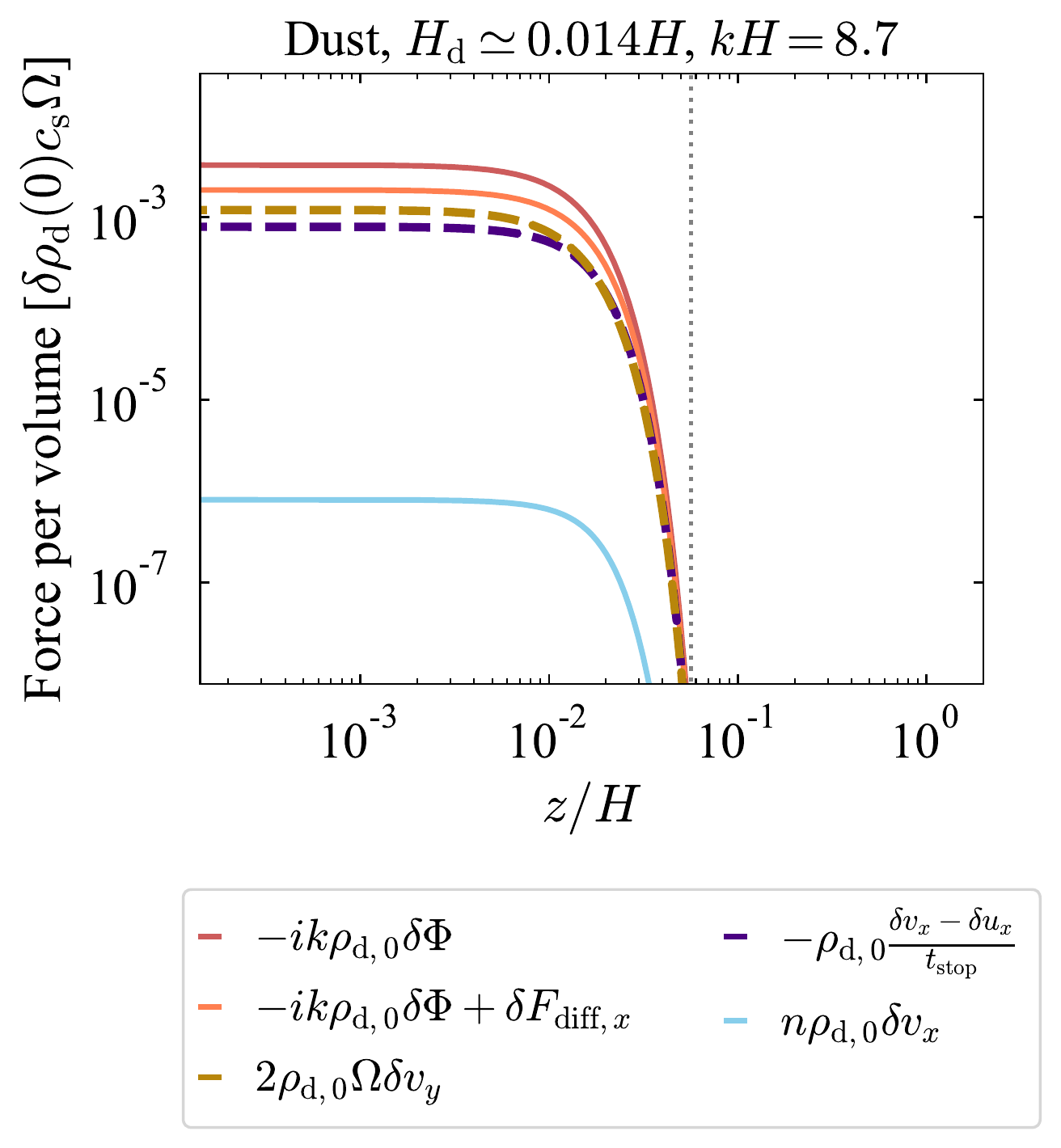} %2 column
%	\hspace{0pt}\raisebox{20pt}{
%	\includegraphics[width=0.5\columnwidth]{./disp_Q3d1.5_alpr1e-4_alpz1e-4_taus05_dgmid05_Hs2Hg_dust_radial_Force.pdf} %1 column
	}
	\end{center}
	\vspace{-30pt}
\caption{The radial force on dust and its balance at $x=-\lambda/4$ in the case 1. As in Figure \ref{fig:eigen_func_fiducial}, a solid (dashed) line represents a region where a physical value is positive (negative). The vertical gray dotted line marks $z=\zsd=4\hd$.} 
\label{fig:dust_radial_force}
\end{figure}

In this section, we explain what kind of force drives dust- and gas-flow, respectively. The radial motion of dust is driven by the self-gravity. Figure \ref{fig:dust_radial_force} shows the force balance of dust at $x=-\lambda/4$, where $\lambda$ is the wavelength $\lambda\equiv 2\pi/k$ (see also Figure \ref{fig:flow_struct_fiducial}). As in the other figures, we use solid and dashed lines to distinguish the sign of the physical values. The red line represents the radial self-gravity, $-ik\rhodup\delta\Phi$, and the orange line shows the sum of the self-gravity and the perturbed pressure-like term representing the effect of the radial diffusion,
\begin{equation}
\delta \fdiffx\equiv -\rhodup\times\delta\left(\frac{D_x\rhog}{\tstop\rhod}\frac{\partial\rhod/\rhog}{\partial x}\right).
\end{equation}
The term $\delta\fdiffx$ is in the direction opposite to the self-gravity. The red and orange lines show that the self-gravity dominates over the radial diffusion. The dark yellow line and the purple line represent the Coriolis force and the gas drag force, respectively. These forces are in the opposite direction compared to the self-gravity (the red line), and the sum of the Coriolis force and the drag force is marginally balanced with the sum of the self-gravity and the pressure-like term (the orange line). The sum of these four forces is equal to $n\rhodup\delta v_x$ (the light blue line), which is in the same direction as the self-gravity (the red line).  The self-gravity slightly dominates over the sum of the other forces, driving the radial concentration of dust. This force balance is the same as what we obtained in the 1D linear analysis \citep[e.g.,][]{Takahashi2014,Tominaga2019}. Therefore, the dust concentration process due to secular GI in the present stratified-disk case can be understood in the same way as in the previous 1D case.

\begin{figure}[t!]%[htp] or [H]
	\begin{center}
	\hspace{100pt}\raisebox{20pt}{
	\includegraphics[width=0.9\columnwidth]{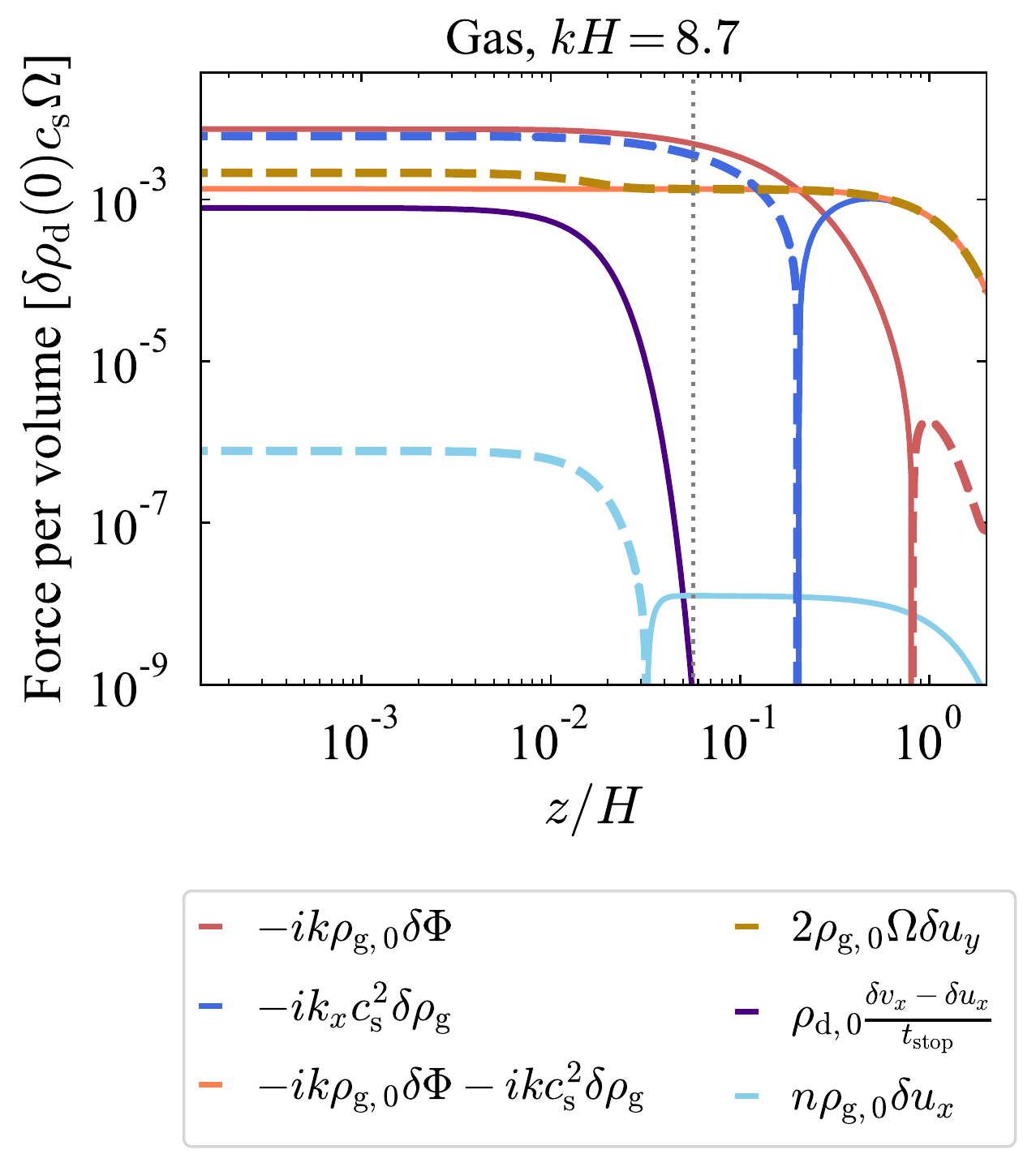} %2 column
%	\hspace{0pt}\raisebox{20pt}{
%	\includegraphics[width=0.5\columnwidth]{./disp_Q3d1.5_alpr1e-4_alpz1e-4_taus05_dgmid05_Hs2Hg_gas_radial_Force2.pdf} %1 column
	}
	\end{center}
	\vspace{-30pt}
\caption{The radial force on gas and its balance at $x=-\lambda/4$ in the case 1. As in Figure \ref{fig:eigen_func_fiducial}, a solid (dashed) line represents a region where a physical value is positive (negative). The vertical gray dotted line marks $z=\zsd=4\hd$.} 
\label{fig:gas_radial_force}
\end{figure}

Next, we explain the driving force of the gas flow. Figure \ref{fig:gas_radial_force} shows the radial force on gas at $x=-\lambda/4$ as a function of the vertical distance $z$. The self-gravity and the pressure gradient force have larger magnitudes in the dust layer than the Coriolis force and the drag force. However, these forces are in the opposite direction to each other with similar magnitudes in the dust layer. As a result, the sum of the self-gravity and the pressure gradient force is much smaller than each of them (the orange line). In the dust layer, especially at $z\lesssim0.01H$, the angular momentum transport from dust to gas tends to cause the radial gas motion in the opposite direction to the dust motion (the light blue line; see also Figure \ref{fig:dust_radial_force}). This can be seen as the enhanced Coriolis force at $z\lesssim0.01H$ (the dark yellow line), which is directed away from the dust density maximum. This enhanced Coriolis force dominates over the sum of the self-gravity and the pressure gradient force (the orange line). The drag force due to the radial dust motion (i.e., the backreaction, the purple line) is in the direction toward the dust density maximum, but this dragging is not enough to stop this radial gas motion. As a result, the gas moves in the opposite direction to the dust motion. 

The Coriolis force and the pressure gradient force dominate over the self-gravity in magnitude at $z\gtrsim0.2H$. The pressure gradient force is the only force in the same direction as the gas radial motion (the blue line) around the surface ($z\gtrsim H$). Thus, the pressure gradient force is the main driver of the radial gas motion around this region. As mentioned above, the gas should behave as if it is incompressible fluid during the linear growth of secular GI. Thus, it is the naive explanation of the gas motion as a whole that the lower-layer gas pushes surrounding gas, triggers the meridional circulation, and leads to the upper-layer gas motion in the opposite direction to the lower-layer gas motion. 

\begin{figure}[t!]%[htp] or [H]
	\begin{center}
	\hspace{100pt}\raisebox{20pt}{
	\includegraphics[width=0.9\columnwidth]{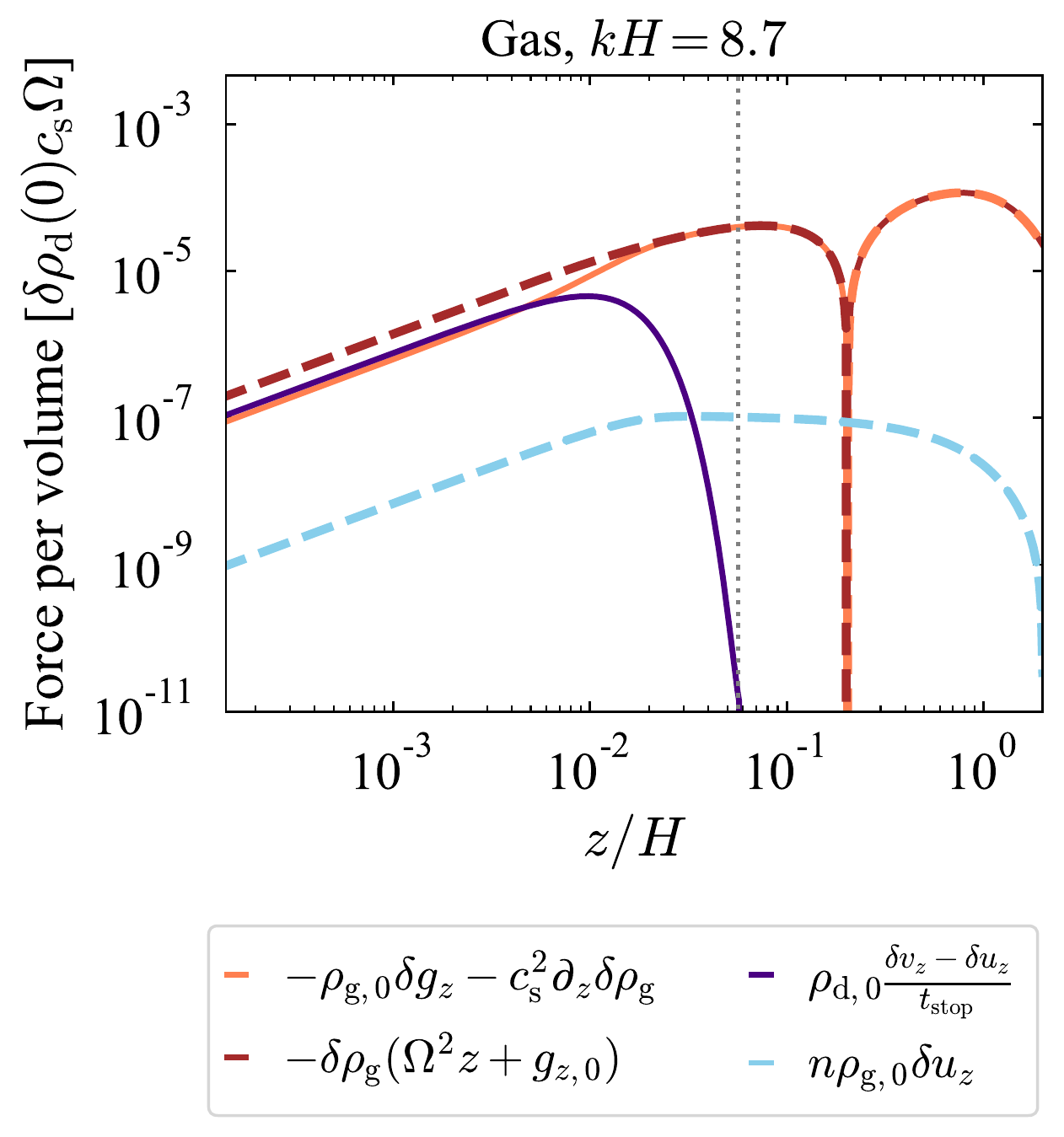} %2 column
%	\hspace{0pt}\raisebox{20pt}{
%	\includegraphics[width=0.5\columnwidth]{./disp_Q3d1.5_alpr1e-4_alpz1e-4_taus05_dgmid05_Hs2Hg_gas_vertical_Force.pdf} %1 column
	}
	\end{center}
	\vspace{-30pt}
\caption{The vertical force on gas and its balance in the case 1. As in Figure \ref{fig:eigen_func_fiducial}, a solid (dashed) line represents a region where a physical value is positive (negative). The vertical gray dotted line marks $z=\zsd=4\hd$. In the legend of the orange line, $\partial_z\delta\rhog$ represents $\partial\delta\rhog/\partial z$.} 
\label{fig:gas_vertical_force}
\end{figure}

\begin{figure*}[t!]%[htp] or [H]
	\begin{center}
	\hspace{0pt}\raisebox{20pt}{
	\includegraphics[width=1.5\columnwidth]{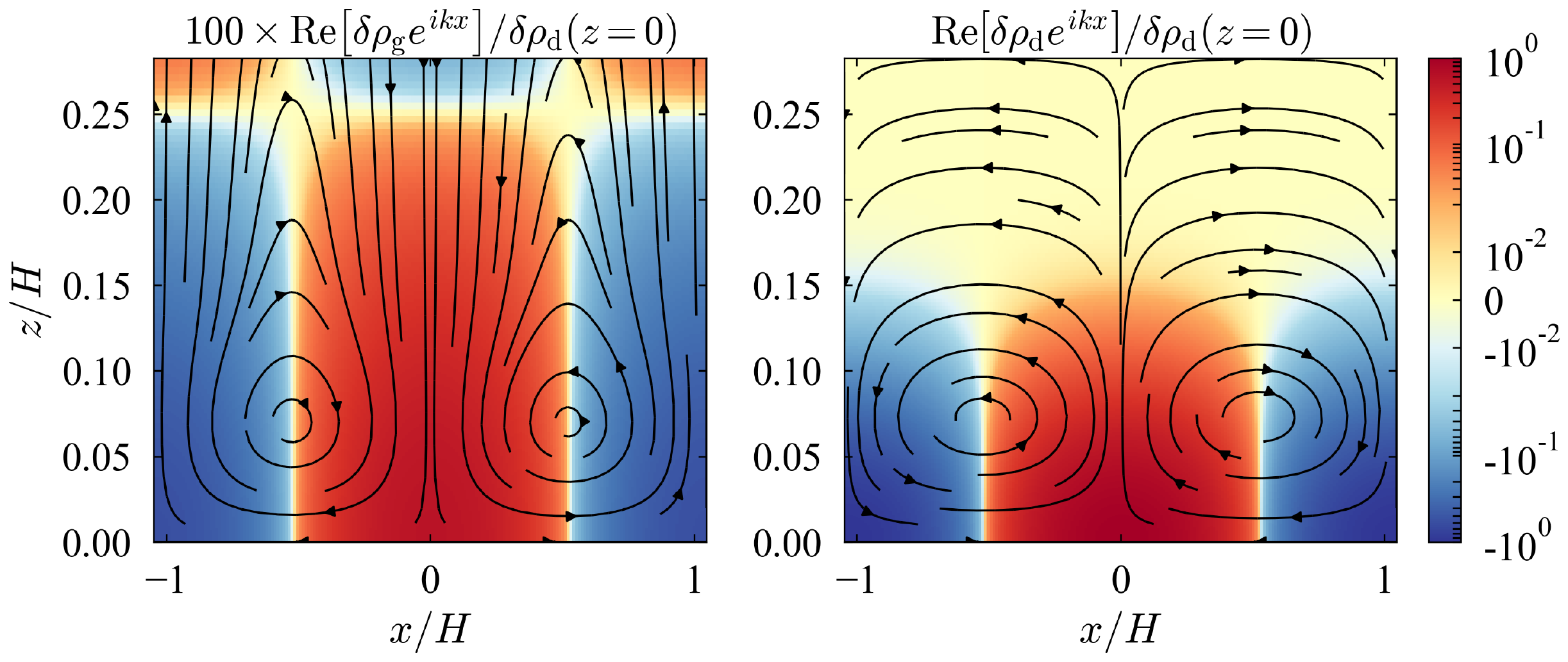} %2 column
%	\hspace{0pt}\raisebox{20pt}{
%	\includegraphics[width=0.9\columnwidth]{./disp_Q3d1.5_alpr3e-4_alpz1e-3_taus02_dgmid03_Hs2Hg_gas_and_dust_density_perturbation.pdf} %1 column
	}
	\end{center}
	\vspace{-30pt}
\caption{Streamlines of gas (the right panel) and dust (the left panel) within the dust layer ($0\leq z\leq \zsd$) in the case 4. The color shows $100\mathrm{Re}\left[\delta\rhog e^{ikx}\right]/\delta\rhod(z=0)$ and $\mathrm{Re}\left[\delta\rhod e^{ikx}\right]/\delta\rhod(z=0)$.} 
\label{fig:flow_struct_2nd_type}
\end{figure*}

Figure \ref{fig:gas_vertical_force} shows the vertical force on gas at $x=0$. The gas vertical velocity (the blue line) is in the direction toward the dust density maximum as shown in Figure \ref{fig:flow_struct_fiducial}. At $z\lesssim 0.2H$, this downward flow is driven not only by the self-gravity but also by the background gravity (the brown line), which is given by
\begin{equation}
-\rhogup\delta\left(\frac{\cs^2}{\rhog}\right)\frac{d\rhogup}{dz} = -\delta\rhog(\Omega^2z+g_{z,0}),
\end{equation}
where we use Equation (\ref{eq:gas_eomz_unp}). We note that the vertical self-gravity is insufficient to solely drive the downward flow around the midplane, and the pressure gradient force dominates over the self-gravity. This is shown by the orange line (see also the black line in the bottom right panel of Figure \ref{fig:eigen_func_fiducial}). The sum of the self-gravity and the pressure gradient force is, however, weaker than the background gravity. In this way, the downward gas motion at $z\lesssim 0.2H$ is mainly driven by the sum of the self-gravity and the background gravity. In the very upper layer ($z\sim H$), the pressure gradient force is the main driver of the downward flow, which can be seen from the fact that the magnitude of $\delta g_z$ is much smaller than what the orange line shows (see the bottom right panel of Figure \ref{fig:eigen_func_fiducial}).

\subsection{Anisotropic diffusion cases}\label{subsec:anistropic_D}
In this subsection, we briefly describe another type of dust flow found in some of the cases with anisotropic diffusion ($D_z\neq D_r$). In the cases with $D_z<D_r$ with $\epsilon(0)\lesssim1$ (see also Table \ref{tab:params_appendix} and Appendix \ref{app:dust_fall}), dust and gas flow structures are similar to those in the previous subsection. In contrast, we find another type of dust flow for $D_z>D_r$, which is prominent in the case 4 in Table \ref{tab:params}. Figure \ref{fig:flow_struct_2nd_type} shows the obtained density perturbations and streamlines of gas and dust for the case 4. As seen in the right panel, the dust also shows the meridional circulation in the dust layer. The direction of this circulation is opposite to the gas circulation, and the dust grains move radially toward the density maximum around the midplane. However, the vertical diffusion is strong in this case, and the dust grains slightly flow out of the dust-rich region. The dust grains flow upward, and then move toward the dust density minimum ($x\simeq \pm H$ in Figure \ref{fig:flow_struct_2nd_type}). 

\begin{figure}[t!]%[htp] or [H]
	\begin{center}
	\hspace{100pt}\raisebox{20pt}{
	\includegraphics[width=0.9\columnwidth]{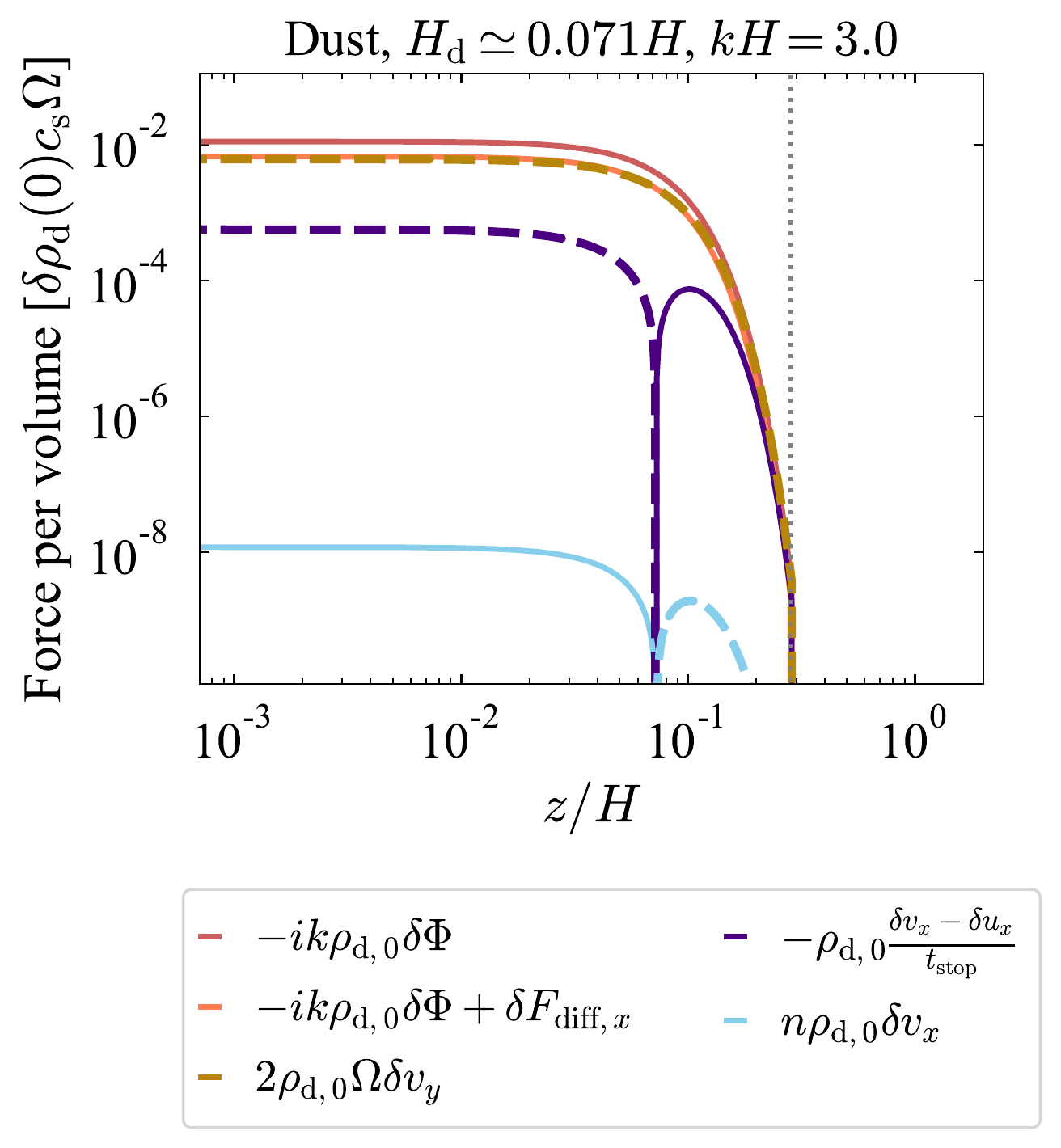} %2 column
%	\hspace{0pt}\raisebox{20pt}{
%	\includegraphics[width=0.5\columnwidth]{./disp_Q3d1.5_alpr3e-4_alpz1e-3_taus02_dgmid03_Hs2Hg_dust_radial_Force.pdf} %1 column
	}
	\end{center}
	\vspace{-30pt}
\caption{The radial force on dust and its balance at $x=-\lambda/4$ in the case 4. As in Figure \ref{fig:eigen_func_fiducial}, a solid (dashed) line represents a region where a physical value is positive (negative). The vertical gray dotted line marks $z=\zsd=4\hd$.} 
\label{fig:dust_radial_force_2nd_type}
\end{figure}

Figure \ref{fig:dust_radial_force_2nd_type} shows the radial force on dust grains. We can see that the self-gravity dominates over $\delta \fdiffx$ as in the case 1 in the region where dust grains move radially toward the density minimum ($z\gtrsim0.1H>\hd$). The Coriolis force, however, dominates over the sum of the self-gravity and $\delta\fdiffx$ in such a region. 

\begin{figure}[t!]%[htp] or [H]
	\begin{center}
	\hspace{100pt}\raisebox{20pt}{
	\includegraphics[width=0.9\columnwidth]{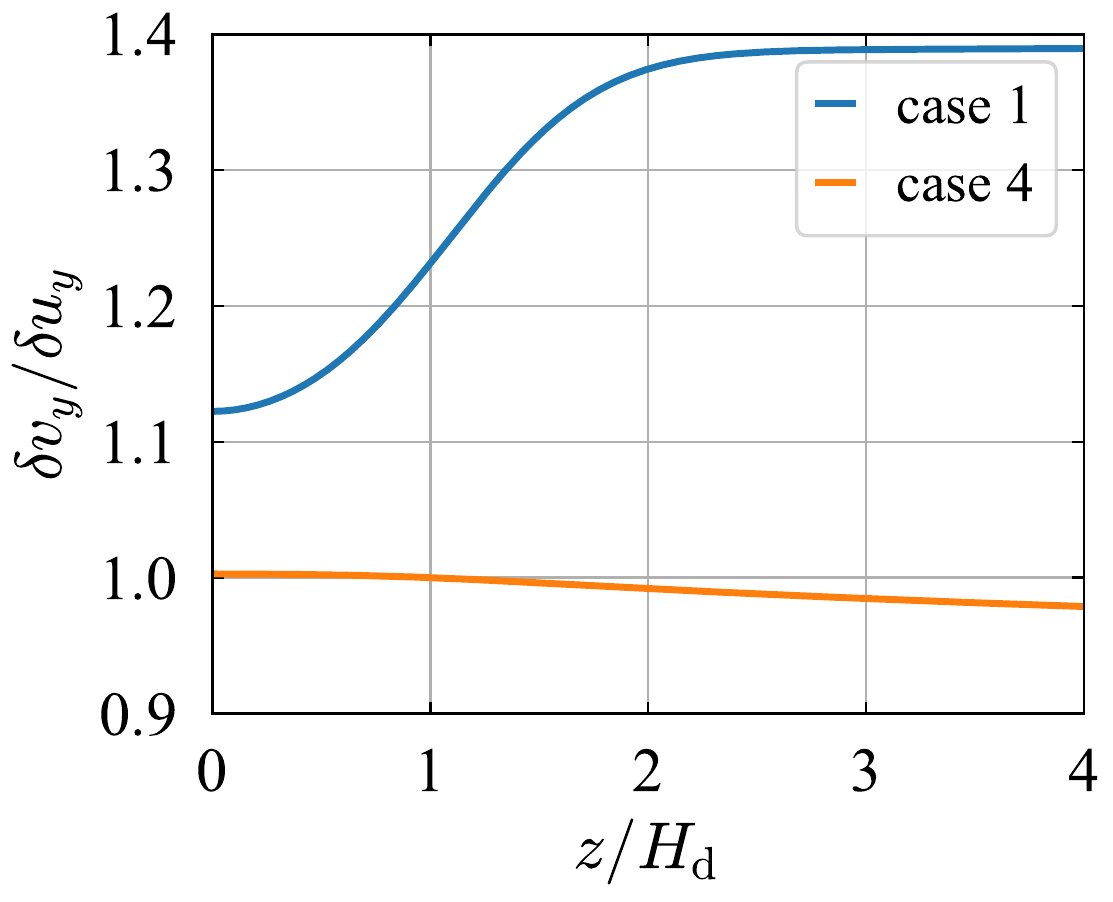} %2 column
%	\hspace{0pt}\raisebox{20pt}{
%	\includegraphics[width=0.5\columnwidth]{./disp_Q3d1.5_alpr1e-4_alpz1e-4_taus05_dgmid05_Hs2Hg__VS__disp_Q3d1.5_alpr3e-4_alpz1e-3_taus02_dgmid03_Hs2Hg_COMPARE_VY-UY.pdf} %1 column
	}
	\end{center}
	\vspace{-30pt}
\caption{Ratio of $\delta v_y$ and $\delta u_y$ for the case 1 (blue line) and for the case 4 (the orange line). The horizontal axis is the vertical distance normalized by the dust scale height. In the case 1, $\delta v_y$ is always larger than $\delta u_y$. In the case 4, however, $\delta v_y$ is smaller at $z>\hd $ than $\delta u_y$. } 
\label{fig:compare_dvy_and_duy_case1_case4}
\end{figure}

The strong Coriolis force on dust in the upper region is due to the relatively large azimuthal velocity of the gas in the upper region. Figure \ref{fig:compare_dvy_and_duy_case1_case4} shows $\delta v_y/\delta u_y$ as a function of $z/\hd$ for the case 1 (the blue line) and the case 4 (the orange line). The azimuthal velocity perturbation of dust is always larger than that of gas in the case 1. This means that the dust grains lose or gain angular momentum so that they can move toward the density maximum. On the other hand, the case 4 shows that $\delta v_y$ is smaller than $\delta u_y$ in the upper region ($z\gtrsim \hd$). This means that the dust grains at $x>0$ ($x<0$) in Figure \ref{fig:flow_struct_2nd_type} get accelerated (decelerated) azimuthally in this upper region, which enhances the radial Coriolis force on the dust grains ($2\rhodup\Omega\delta v_y$). As a result, the dust grains move away from the density maximum ($x=0$). In this way, we attribute the backward dust motion at $z\gtrsim \hd$ to the sum of the enhanced Coriolis force and the diffusion.

This kind of dust flow structure does not appear in the previous 1D analysis. It should be noted that the 1D secular GI is stable for the parameters of the case 4 (see Table \ref{tab:params}). This highlights the importance to consider the vertical structure and motion even when a dust disk is much thinner than a gas disk.

\subsection{Condition for stratified secular GI}\label{subsec:cond}

Finally, we describe the growth condition of secular GI in vertically stratified disks. We conduct a parameter survey to empirically obtain the growth condition. Specifically, we change the midplane dust-to-gas ratio $\epsilon(0)$ and $D_z/D_r$ to see the impact of the vertical thickness of the dust. The list of parameters is presented in Appendix \ref{app:params_cond}. We find that the stability is well described by the total dust-to-gas ratio $\varepsilon=\sigmad/\sigmag$ and the Toomre's $Q$ for the dust,
\begin{equation}
Q_{\dst}(\tilde{D}_r)\equiv\frac{\sqrt{D_r/\tstop}\Omega}{\pi G\sigmad}=\frac{\sqrt{\tilde{D}_r/\taus}\cs\Omega}{\pi G\sigmad}.\label{eq:Q_dust}
\end{equation}

\begin{figure}[t!]%[htp] or [H]
	\begin{center}
	\hspace{100pt}\raisebox{20pt}{
	\includegraphics[width=0.9\columnwidth]{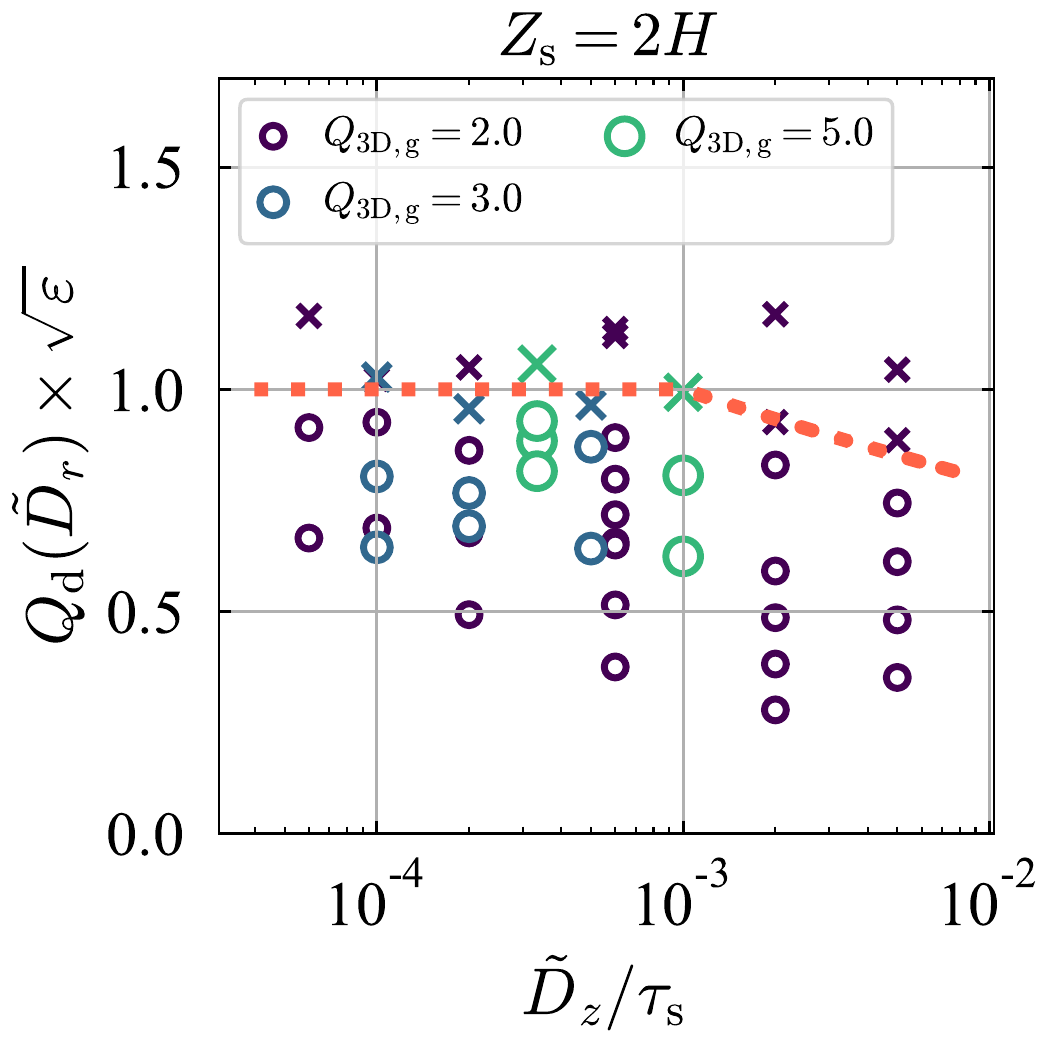} %2 column
%	\hspace{0pt}\raisebox{20pt}{
%	\includegraphics[width=0.5\columnwidth]{./NEW_parame_study_with_growth_rate_criteria_nct_1e-9.pdf} %1 column
	}
	\end{center}
	\vspace{-30pt}
\caption{Results of the parameter studies focusing on the stability of secular GI. The open circles represent the cases where secular GI operates while the cross marks represent the stable cases. The orange broken dashed line is a rough boundary between the stable and unstable regions (Equation (\ref{eq:rough_cond})). We find that the growth condition is clearly described in terms of $Q_{\dst}$ and the total dust-to-gas ratio $\varepsilon=\sigmad/\sigmag$.} 
\label{fig:growth_condition}
\end{figure}

Figure \ref{fig:growth_condition} shows the result. The vertical axis is $Q_{\dst}(\tilde{D}_r)\sqrt{\varepsilon}$, and the horizontal axis is $\tilde{D}_z/\taus$, which represents the vertical thickness of the dust disk relative to the gas thickness ($\hd/H=(\tilde{D}_z/\taus)^{0.5}$). The open circles represent the cases where secular GI can grow. The cross marks in the figure represent the cases where secular GI is stable. The orange broken dashed line is 
\begin{equation}
Q_{\dst}(\tilde{D}_r)\sqrt{\varepsilon}=\min\left[1, \left(\frac{\tilde{D}_z/\taus}{10^{-3}}\right)^{-0.1}\right],\label{eq:rough_cond}
\end{equation}
which roughly traces the unstable-stable boundary. The dependence on the vertical thickness is very weak although the critical value of $Q_{\dst}(\tilde{D}_r)\sqrt{\varepsilon}$ slightly decreases as the vertical thickness increases. Thus, we may neglect the $\hd$-dependence and provide a rough condition for secular GI as follows:
\begin{equation}
Q_{\dst}\sqrt{\frac{\sigmad}{\sigmag}}\lesssim 1.\label{eq:rough_cond_2}
\end{equation}

In our previous analysis based on the thin-disk model \citep[][]{Takahashi2014,Tominaga2019}, the growth condition is given by the following equation
\begin{equation}
Q < \sqrt{\frac{\varepsilon(1+\varepsilon)\tstop\cs^2}{D_r}},\label{eq:RT19_eq33}
\end{equation}
where we assume $(1+\varepsilon)D_r\ll \varepsilon\cs^2\tstop$ \citep[see Equation (33) of][]{Tominaga2019}. One can rearrange the left-hand side as
\begin{equation}
Q=\frac{\cs\Omega}{\pi G\sigmad}\times\varepsilon,
\end{equation}
and then obtain
\begin{equation}
\frac{\sqrt{D_r/\tstop}\Omega}{\pi G\sigmad}\sqrt{\varepsilon} < \sqrt{1+\varepsilon}\sim 1.
\end{equation}
This condition obtained from the 1D analysis is almost equivalent to the condition obtained in the present 3D analysis (Equation (\ref{eq:rough_cond_2})). Thus, one can still use the previous condition (\ref{eq:RT19_eq33}) to discuss the stability of a disk against secular GI. The above result provides important insight into the 1D growth condition: 
\begin{enumerate}
\item the total dust-to-gas surface density ratio $\varepsilon=\sigmad/\sigmag$ is one ingredient to determine the stability,
\item the Toomre's $Q$ value for gas in Equation (\ref{eq:RT19_eq33}) should be estimated with the total gas surface density and not with the gas surface density within the dust sublayer,
\item the radial diffusivity is more important for the growth condition than the vertical diffusivity.
\end{enumerate}

If one uses the gas surface density within the dust sublayer, $\Sigma_{\gas,\mathrm{sub}}\simeq\sigmag\hd/H=\sigmag\sqrt{D/\tstop\cs^2}$ with the assumption of isotropic diffusion, $D_r=D_z=D$, and replace $Q$ and $\varepsilon$ in Equation (\ref{eq:RT19_eq33}) as
\begin{equation}
Q\to\frac{\cs\Omega}{\pi G\Sigma_{\gas,\mathrm{sub}}},
\end{equation}
\begin{equation}
\varepsilon\to\frac{\sigmad}{\Sigma_{\gas,\mathrm{sub}}},
\end{equation}
one then arrives at the following equation \citep[see also][]{Lesur2022}
\begin{equation}
Q_{\dst}\sqrt{\varepsilon} < \left[\sqrt{\frac{D}{\tstop\cs^2}}+\varepsilon\right]^{1/2}.\label{eq:sublayer_cond}
\end{equation}
This cannot explain the growth condition obtained in this work. Moreover, Equation (\ref{eq:sublayer_cond}) significantly underestimates the viability of secular GI since the right-hand side is much less than unity for $D/\tstop\cs^2<10^{-2}$ and $\varepsilon\sim10^{-2}$.

In summary, we propose Equation (\ref{eq:rough_cond_2}) as the rough growth condition for secular GI in weakly turbulent disks.

\section{Discussions}\label{sec:discussion}
\subsection{Dependence on the outer boundary location}\label{subsec:zs_dependence}
In the main analyses of this paper, we set the outer boundary of gas to be $\zs=2H$. To discuss the $\zs$-dependence of the growth rates, we linearly space the domain with sufficiently high resolution of 20000 grids$/H$ and derive the growth rates for a given $\zs$.

\begin{figure}[t!]%[htp] or [H]
	\begin{center}
	\hspace{100pt}\raisebox{20pt}{
	\includegraphics[width=0.9\columnwidth]{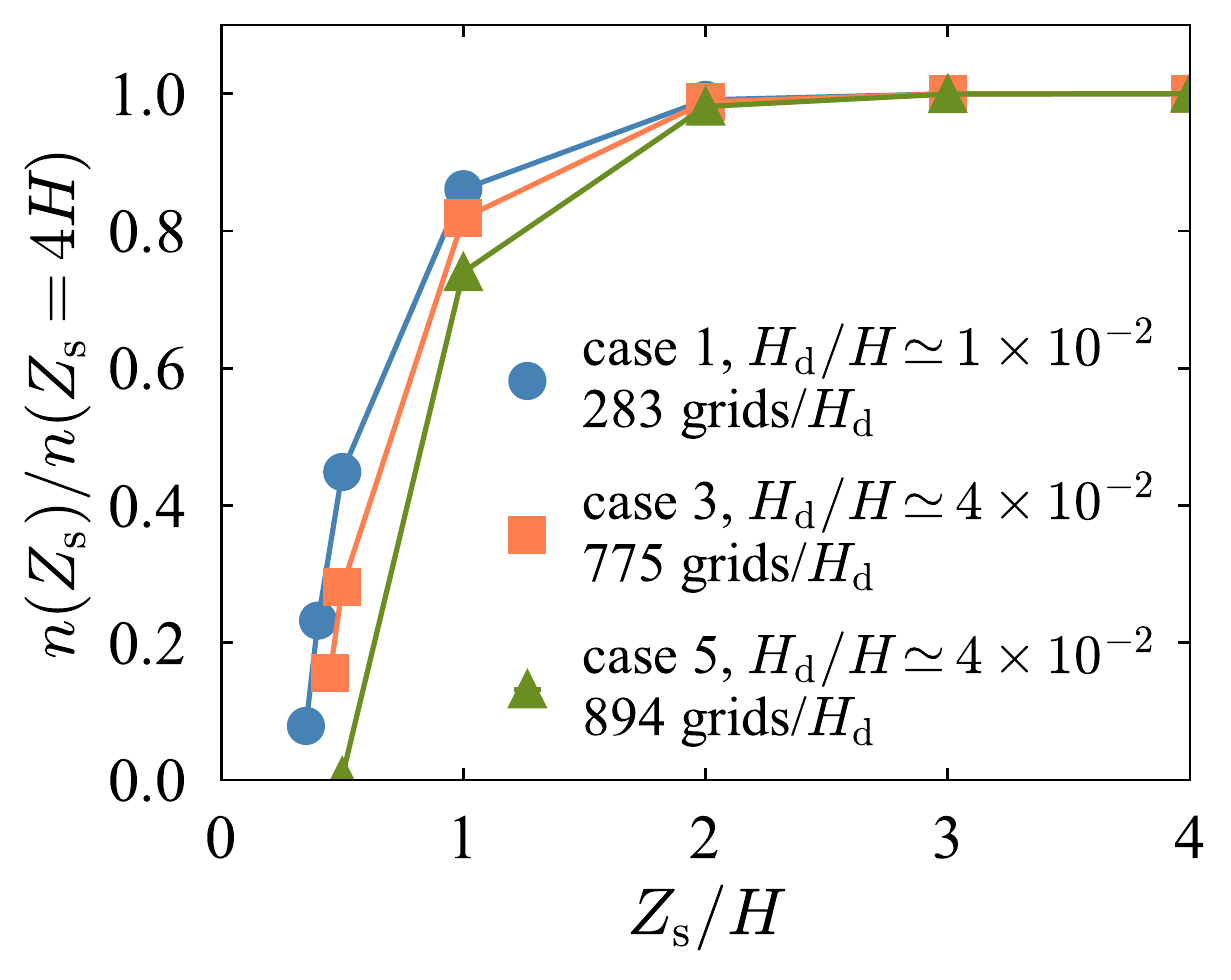} %2 column
%	\hspace{0pt}\raisebox{20pt}{
%	\includegraphics[width=0.5\columnwidth]{./Growth-rate_convergence_for_paper.pdf} %1 column
	}
	\end{center}
	\vspace{-30pt}
\caption{$\zs$-dependence of the growth rates in the cases 1, 3, and 5. The vertical axis is normalized by the growth rate for $\zs=4H$. The resolution in space is 283 grids, 775 grids, and 894 grids per $\hd$ in each case. We find that the location of the outer gas boundary substantially affects the secular GI regardless of the fact that the aerodynamical interaction important for secular GI is operational only within the thin dust layer.} 
\label{fig:zs_dependence}
\end{figure}

Figure \ref{fig:zs_dependence} shows the obtained $\zs$-dependence of the growth rates in the cases 1, 3, and 5 in Table \ref{tab:params}. The resolution is 283 grids$/\hd$, 775 grids$/\hd$, and 894 grids$/\hd$, respectively, with which we resolve the very thin dust layer ($\hd/H < 0.1$). We can see that the growth rates decrease as we move the outer gas boundary toward the midplane. The growth rate decreases by a factor of $\sim5-10$ from $n(\zs=4H)$ even for $\zs\sim0.5H\gg\hd$. This is one noticeable finding of this work. Although the dust concentration driving secular GI takes place only within the dust layer, the gas motion well above the dust layer is important for secular GI to operate. When we adopt lower $\zs$, we regard the upper gas as the steady source of the external pressure to the lower layer (see Equation (\ref{eq:freeBC_gas})). Figure \ref{fig:zs_dependence} thus means that the non-steady upper gas motion is important for secular GI. It should be also noted that the growth rate converges enough for $\zs\geq2H$, and thus this validates our choice of $\zs$ in Section \ref{sec:linear_prob}. 

We speculate that we may need to deal with such a large-scale gas circulation in numerical simulation of secular GI as well as dust concentration around the midplane. Secular GI would be unphysically stabilized if one adopts a small vertical domain size (e.g., $0.1H$). We will address this point by conducting multidimensional simulations in future work.

\subsection{Comparison with observations: Dust mass budget}\label{subsec:comp_obs}
As presented in the previous section, Equation (\ref{eq:rough_cond_2}) provides the growth condition of secular GI. We can rearrange this formula to a condition on the dust disk mass, which can be estimated from continuum observations. From Equation (\ref{eq:Q_dust}), we can write the Toomre's $Q$ value for dust as follows:
\begin{equation}
Q_{\dst}(\tilde{D}_r)=\frac{\sqrt{\tilde{D}_r/\taus}M_{\ast}H/R}{\pi R^2\sigmad}.
\end{equation}
We then rearrange Equation (\ref{eq:rough_cond_2}) as follows
\begin{align}
\pi R^2\sigmad &\gtrsim \sqrt{\frac{\sigmad}{\sigmag}}M_{\ast}\frac{H}{R}\sqrt{\frac{\tilde{D}_r}{\taus}}\notag\\
&\simeq 3\times 10^{-4}M_{\ast}\left(\frac{\sigmad/\sigmag}{10^{-2}}\right)^{1/2}\left(\frac{H/R}{10^{-1}}\right)\left(\frac{\tilde{D}_r/\taus}{10^{-3}}\right)^{1/2}.\label{eq:rough_Md_cond}
\end{align}
Although the physical values in this condition are local values at a certain radius, the left-hand side roughly represents the dust disk mass (the enclosed mass). We then regard the right-hand side as the critical dust disk mass required for secular GI. One should note that the terms on both sides of Equation (\ref{eq:rough_Md_cond}) depend on $\sigmad$, and the right-hand side has weaker dependence on $\sigmad$. Thus, for a given gas surface density, larger dust masses will trigger secular GI more easily.

From the observational point of view, the dust mass is easier to estimate than the gas mass since the most abundant hydrogen molecule is too dark at low disk temperature for lack of a dipole moment. Even if the dust thermal emission is optically thick, we may estimate the lower limit of the dust mass although there are uncertainties in dust opacity \citep[e.g., see review by][and reference therein]{Andrews2020}. Thus, once the dust mass is observationally constrained (the left-hand side of Equation (\ref{eq:rough_Md_cond})), we can discuss the viability of secular GI in the observed disks based on the total dust-to-gas ratio ($\sigmad/\sigmag$), and the other factors on the right-hand side. Interestingly, the critical dust disk mass is proportional to $(\sigmad/\sigmag)^{1/2}$: the critical dust mass is lower if the disk is richer in gas. Such a dust-depleted disk can be expected when dust grains grow enough and drift inward \citep[e.g.,][]{Brauer2008}. Since the stopping time is expected to be large ($\taus\sim 0.1$) in this case, the critical dust mass will become lower than for small dust, and secular GI can operate more easily.

\begin{figure}[t!]%[htp] or [H]
	\begin{center}
	\hspace{100pt}\raisebox{20pt}{
	\includegraphics[width=0.8\columnwidth]{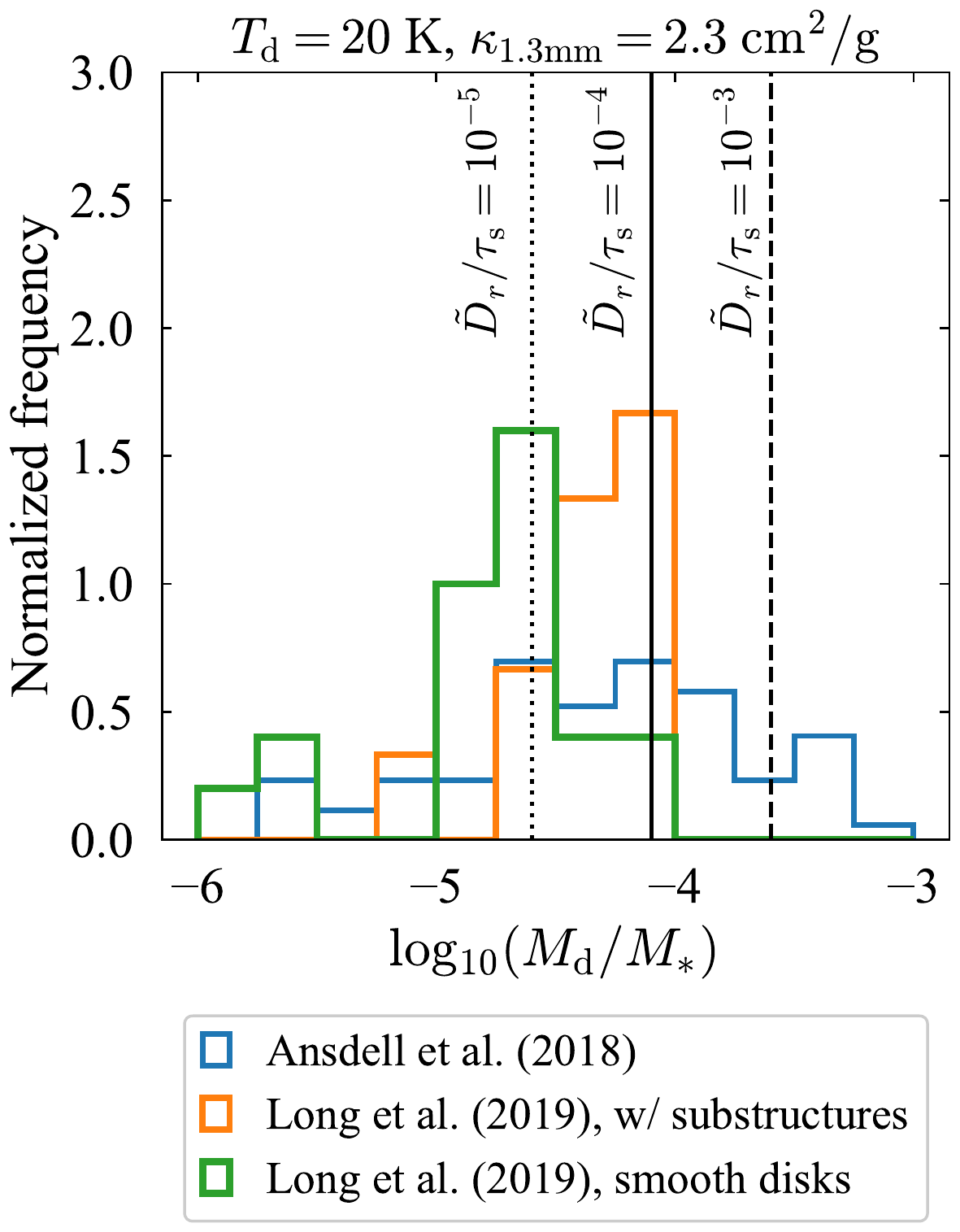} %2 column
%	\hspace{0pt}\raisebox{20pt}{
%	\includegraphics[width=0.5\columnwidth]{./stratified_SGI_cond_kappa2.25_Tave20.0_hist.pdf} %1 column
	}
	\end{center}
	\vspace{-30pt}
\caption{Comparison between the critical mass ratio between the dust disk and the central star (Equation (\ref{eq:rough_Md_cond}), black lines) and the observationally estimated mass ratios reported in \citet{Ansdell2018} (see their Table 1 and the references therein) and those based on \citet{Long2019} (see their Tables 1 and 3, and the references therein). To plot the critical $M_{\dst}/M_{\ast}$, we assume $\sigmad/\sigmag=10^{-2}$ and $H/R=10^{-1}$ and adopt $\tilde{D}_r/\taus=10^{-5},\;10^{-4},\;10^{-3}$. We assume the emission to be optically thin to estimate the dust disk masses (Equation (\ref{eq:opt_thin})). As for the samples in \citet{Ansdell2018}, we use its subsamples whose reported continuum flux is positive. The green histogram includes smooth disks in both single star systems and binaries/multiple systems. All histograms are normalized so that the area of the histogram is equal to unity.   } 
\label{fig:obs_comparison}
\end{figure}

In the following, we adopt the canonical value of $10^{-2}$ for dust-to-gas ratio for conservative discussion. For a stellar mass of $0.1\msun-1\msun$, the required dust mass is $\sim10\merth-100\merth$. We compare the critical dust mass for secular GI with observationally estimated dust masses. If one assumes the optically thin emission, one can estimate the dust disk mass from continuum flux, $F_{\nu}$:
\begin{equation}
M_{\dst} =  \frac{d^2F_{\nu}}{ \kappa_{\nu}B_{\nu}(T_{\dst})},\label{eq:opt_thin}
\end{equation}
where $\kappa_{\nu}$ is the absorption opacity at frequency $\nu$, $B_{\nu}(T_{\dst})$ is the Plank function at dust temperature $T_{\dst}$,  and $d$ is the distance to a disk \citep[e.g.,][]{Hildebrand1983,Beckwith1990}. This formula has been frequently used in observational studies for statistical analyses of the disk masses \citep[e.g.,][]{Andrews2005,Andrews2013,Ansdell2016,Ansdell2018,Long2019}. Figure \ref{fig:obs_comparison} shows the normalized frequency of $M_{\dst}/M_{\ast}$ estimated from 1.3 mm continuum observations utilizing the Atacama Large Millimeter/submillimeter Array (ALMA) in \citet{Ansdell2018} and \citet{Long2019} (see the references therein for the stellar masses and the distances for the individual stars). 
We adopt the dust temperature and opacity adopted in \citet{Ansdell2016,Ansdell2018}\footnote{The opacity in \citet{Ansdell2016,Ansdell2018} is assumed to be $10\;\mathrm{cm^2/g}$ at $10^3$ GHz, which gives $\simeq2.3\;\mathrm{cm^2/g}$ at 225 GHz.} to estimate $M_{\dst}/M_{\ast}$ from the data in \citet{Long2019}:
 $T_{\dst}=20$ K and $\kappa_{1.3\;\mathrm{mm}}=2.3\;\mathrm{cm}^2\mathrm{/g}$. We also plot the critical values of $M_{\dst}/M_{\ast}$ for secular GI to operate assuming $\sigmad/\sigmag=10^{-2}$ and $H/R=10^{-1}$ (Equation (\ref{eq:rough_Md_cond})). If disk turbulence is so weak that $\tilde{D}_r/\taus$ is $\sim 10^{-5}-10^{-4}$, the higher-mass disks among the samples seem to have enough dust masses to trigger secular GI. It should be noted that the dust masses in those samples would be higher and would be more unstable to secular GI if the disks are optically thick.

\citet{Tobin2020b} conducted statistical characterization of Class 0 and Class I disks with ALMA and the Very Large Array (VLA) and compared the dust masses among Class 0-II disks (see Figures 14 and 15 therein). They found that the dust mass is higher in younger disks. Thus, we expect that secular GI is easier to operate in younger disks as suggested in the previous studies \citep[e.g.,][]{Takahashi2014,Takahashi2016}.

As mentioned above, the critical dust disk mass depends on the radial diffusivity of dust and the dust size through $\tilde{D}_r/\taus$ (the third parenthesis on the right-hand side of Equation (\ref{eq:rough_Md_cond})). If gas turbulence is nearly isotropic, we can naively replace $\tilde{D}_r/\taus$ with $\tilde{D}_z/\taus$, which we can estimate observationally from dust scale heights \citep[][]{Pinte2016,Doi2021}. \citet{Pinte2016} estimated the strength of turbulence in the HL Tau disk and found $\alpha\simeq3\times10^{-4}$, with which secular GI can operate for mm-sized dust grains \citep[][]{Takahashi2016}. \citet{Doi2021} presented a method to estimate a dust scale height based on intensity variation in the azimuthal direction. They constrained the dust scale height at the inner and outer rings of HD 163296 and found $\alpha/\taus>2.4$ at the inner ring and $\alpha/\taus < 1.1\times 10^{-2}$ at the outer ring. This indicates that the outer ring is preferable for secular GI as they noted.

If gas turbulence is substantially anisotropic, we cannot replace $\tilde{D}_r/\taus$ with $\tilde{D}_z/\taus$ in Equation (\ref{eq:rough_Md_cond}), and we have to observationally estimate the radial diffusivity to discuss the viability of secular GI. Assuming the existence of pressure bumps and the equilibrium between dust diffusion and dust accumulation toward the pressure bump, \citet{Dullemond2018} estimated $\alpha/\taus$ at some rings observed in the Disk Substructures at High Angular Resolution Project \citep[DSHARP,][]{Andrews2018}. They found $\alpha/\taus\sim10^{-2}-10^{-1}$ and even larger values $\sim1$ (see Table 3 therein). These values are so large that secular GI may be hardly operational in the rings. However, it should be noted that there are some effects widening ring widths: strong backreaction from dust to gas \citep[][]{Kanagawa2018} and wakes from a gap-opening planet \citep[][]{Bi2022}. If those processes operate in the DSHARP disks, the turbulence strength might be lower than estimated by \citet{Dullemond2018}, which is preferable for secular GI. Secular GI can be also a mechanism to form the observed rings \citep[][]{Takahashi2014,Takahashi2016}. In such a case, we cannot adopt the value of $\alpha/\taus$ in \citet{Dullemond2018} since they considered the dust trapping in a pressure bump as the cause of the rings. For conclusive discussion on the viability of secular GI, we need more comprehensive modeling and observational estimation of the radial diffusivity and $\tilde{D}_r/\taus$, which is beyond the scope of this paper.

\subsection{Neglected processes in this work}\label{subsec:visc_shear}
In this work, we neglected the drift motion of dust and turbulent viscosity in order to focus on the effect of the vertical structure on secular GI. Here, we give brief comments on those effects. 

The previous studies found that the dust drift does not affect the growth rate of secular GI \citep[][]{Youdin2005a,Tominaga2020}. Dust density perturbations just propagate inward at the drift speed in the presence of the background drift motion. We naively speculate that this is the case even with the vertical stratification. The important driving force of secular GI is self-gravity. The present analysis shows that the main source of the self-gravity perturbation is the dust density perturbation confined in the dust layer ($\delta\rhod\gg\delta\rhog$; see Figure \ref{fig:eigen_func_fiducial}). The dust density perturbation will drift inward uniformly for a given dust size unless the dust layer is so thick that the drift speed changes vertically because of the change in the gas density. In such a case, the drift motion would not change the growth rate of secular GI as in the previous razor-thin-disk analysis. We need to conduct multidimensional analyses with the drift to understand how the midplane drift motion affects the large-scale meridional circulation of gas.

The dust-gas drift motion introduces streaming instability \citep[e.g.,][]{YoudinGoodman2005}. Besides, \citet{Lin2021} found that vertical shear induced by dust triggers another type of streaming instability called vertically shearing streaming instability \citep[VSSI, see also][]{Ishitsu2009}. Those instabilities operate at much shorter wavelengths than secular GI, and thus we may expect that secular GI can coexist with those streaming instabilities. In our future work, we will analyze those modes comprehensively and investigate potential interaction among them.

According to the previous study, turbulent gas viscosity augments secular GI, and even introduces another secular instability named TVGI \citep{Tominaga2019}. In contrast to the previous analyses with the razor-thin disk model, the viscous momentum transport occurs in both radial and vertical directions. It is known that gravitational instability of a dusty-gas disk is augmented by the vertical momentum transport \citep[e.g.,][]{Ziglina2016,Makalkin2017,Makalkin2018}. This might indicate that vertical momentum transport due to gas viscosity further augments secular GI. We will investigate the impact of the viscosity on secular GI as well as the drift motion in future work.

The present analyses include the dust diffusion as a result of the interaction between dust and gas turbulence. The gas turbulence also leads to local dust clustering \citep[e.g.,][]{Cuzzi2001,Pan2011}, which the diffusion model cannot treat rigorously. The spatial scale of clustering is expected to be smaller than the scale of secular GI (e.g., $\sim H$). Thus, we may regard the clustering just as reduction of the diffusivity. The clustering will then lead to faster growth of secular GI. We will study this process in more detail in future work.

\section{Summary}\label{sec:summary}
Secular GI is one of the promising mechanisms for explaining planetesimal formation \citep[e.g.,][]{Ward2000,Youdin2005a,Youdin2005b,Youdin2011,Takahashi2014}. The previous studies utilized only the razor-thin disk model (1D model) in which the vertically stratified structures are neglected. The previously derived condition for secular GI is thus written in terms of the vertically integrated physical values such as the gas and dust surface densities. However, it has been unclear whether or not we can simply use those integrated values to discuss secular GI because the aerodynamical dust-gas interaction only operates within the vertically thin dust layer around the midplane \citep[e.g.,][]{Latter2017,Tominaga2019}. To address this issue, we conducted vertically global linear analyses taking into account the vertical structure and motion of dust and gas. 

We found that the dust motion is almost radial in isotropic turbulence cases while gas shows meridional circulation. The gas motion is nearly incompressible and $\delta\rhog$ is much smaller than $\delta \rhod$ around the midplane. Besides, the gas radial motion is in the direction away from the local maximum of the dust density perturbation, which is opposite to the gas motion in the razor-thin disk model. This meridional gas circulation highlights the importance of considering the vertical motion and structure. We also found that the gas surface density perturbation is in anti-phase with the dust surface density perturbation, which is also in contrast to the previous vertically integrated models. This indicates that secular GI creates dust rings in gaps of the gas surface density although the gas substructure should be very weak ($|\delta\sigmad|\gg|\delta\sigmag|$) as in the previous models \citep[][]{Tominaga2020,Pierens2021}.

The growth rate is only a few times smaller than the growth rate of 1D secular GI with the disk thickness correction in the Poisson equation \citep[][]{Vandervoort1970,Shu1984}. The unstable wavelengths are similar to those in the 1D model. We also found a case where 1D secular GI is stable but secular GI with the vertical motion can operate (the case 4 in Table \ref{tab:params}, see also Figure \ref{fig:flow_struct_2nd_type}), which also highlights the importance of the vertical motion.

We conducted the parameter survey and empirically derived the condition for secular GI to operate in terms of the dust disk mass. Interestingly, the derived condition is similar to the previous 1D condition which is represented not by the gas surface density in the dust layer but by the total gas surface density \citep[][]{Takahashi2014,Tominaga2019}. Therefore, we can safely use the previous 1D growth condition without any modification, especially for weakly turbulent disks.

We also investigated the dependence of the growth rate on the outer boundary of gas ($z=\zs$) and found that the location needs to be above $\simeq 2H$ for convergence of the growth rate. This supports the above conclusion that the total gas surface density should be used in the growth condition. We attribute the $\zs$-dependence of the growth rate to the nearly incompressible motion that is vertically extended toward $\zs$. The growth rates are underestimated if one assumes the upper gas to be steady and regards it just as the source of the external pressure to the lower layer.

As shown in this paper, the gas well above the dust layer plays a role in driving secular GI regardless of the fact that the aerodynamical interaction operates only within the thin dust layer. Numerical simulations that resolve the dust layer and treat the motion of the upper gas will be necessary to reveal nonlinear development of secular GI and physical properties of resulting dust clumps that develop toward planetesimals.

\acknowledgments
We thank Hidekazu Tanaka and Min-Kai Lin for useful discussion. We also thank the anonymous referee for constructive comments. This work was supported by JSPS KAKENHI Grant Nos., 21K20385 (R.T.T.), 16H02160, 18H05436, 18H05437 (S.I.), 19K14764, 18H05438 (S.Z.T.). This work was also supported by NAOJ ALMA Scientific Research Grant Code 2022-20A. R.T.T. is also supported by RIKEN Special Postdoctoral Researchers Program. 


%\software{Mathematica \citep{Wolfram}, Matplotlib \citep{Hunter2007} }


%\newpage

%\newpage
\appendix
\section{Dust-fall mode}\label{app:dust_fall}
Figure \ref{fig:growth_condition} includes cases with $\epsilon(0)> 5$ and $\qthree=5$, for which secular GI can grow (see also Table \ref{tab:params_appendix}). Among them, the case with $D_z<D_x$ shows dust downward flow at the top of the density maximum, which does not appear in the other cases. Although this type of secular GI (``dust-fall mode") is minor among our results, we briefly describe its mode properties for completeness.

\begin{figure*}[t!]%[htp] or [H]
	\begin{center}
	\hspace{0pt}\raisebox{20pt}{
	\includegraphics[width=0.75\columnwidth]{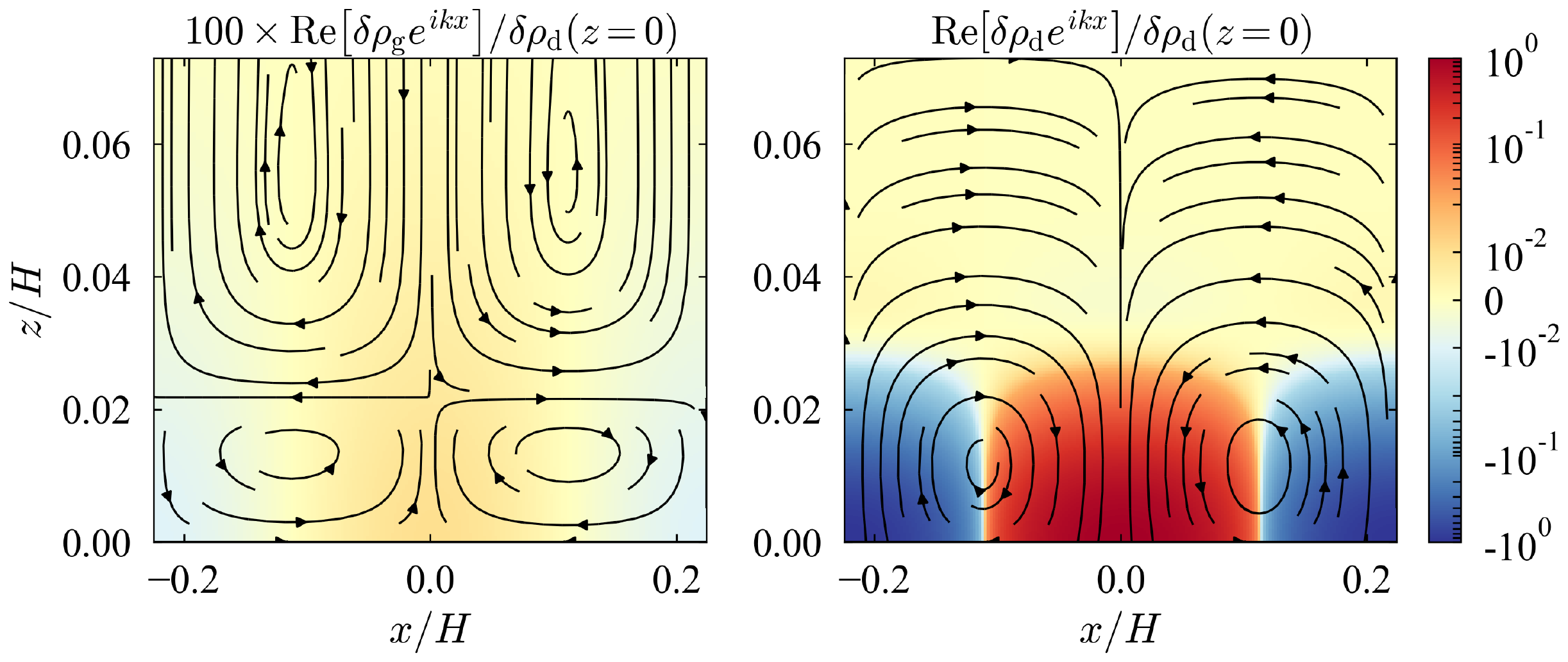} %2 column
%	\hspace{0pt}\raisebox{20pt}{
%	\includegraphics[width=0.9\columnwidth]{./Q3d5_alpr3e-4_alpz1e-4_taus03_dgmid8_Hs2Hg_gas_and_dust_density_perturbation.pdf} %1 column
	}
	\end{center}
	\vspace{-30pt}
\caption{Streamlines of gas (the left panel) and dust (the right panel) at $|z|\leq 4\hd$ of the dust-fall mode with $kH=14$. The physical parameters are $\qthree=5,\;\taus=0.3\;\tilde{D}_r=3\times10^{-4},\;\tilde{D}_z=1\times10^{-4}$ and $\epsilon(0)=8$. The color shows $100\mathrm{Re}\left[\delta\rhog e^{ikx}\right]/\delta\rhod(0)$ and $\mathrm{Re}\left[\delta\rhod e^{ikx}\right]/\delta\rhod(0)$ in the left and right panel, respectively.} 
\label{fig:flow_struct_fall}
\end{figure*}

\begin{figure*}[t!]%[htp] or [H]
	\begin{center}
	\hspace{0pt}\raisebox{20pt}{
	\includegraphics[width=0.75\columnwidth]{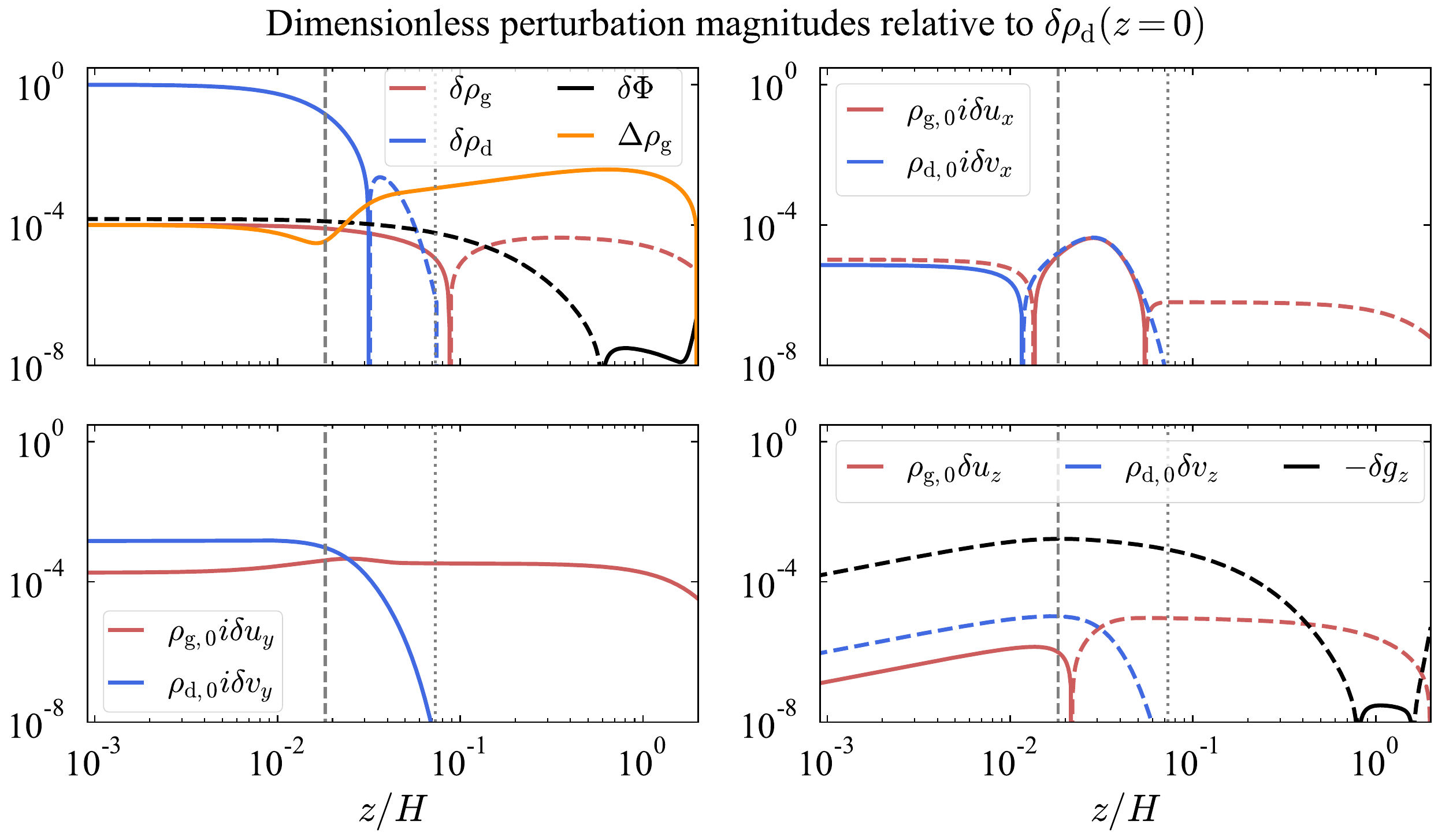} %2 column
%	\hspace{0pt}\raisebox{20pt}{
%	\includegraphics[width=0.9\columnwidth]{./Q3d5_alpr3e-4_alpz1e-4_taus03_dgmid8_Hs2Hg_eigen_func_vertical_log_w_Lag.pdf} %1 column
	}
	\end{center}
	\vspace{-30pt}
\caption{Vertical profiles of the perturbed physical values of the dust-fall mode. The plotted values are normalized by the midplane dust density perturbation $\delta\rhod(0)$, the isothermal sound speed $\cs$, and the Keplerian angular velocity $\Omega$. In each panel, the solid lines represent positive values while the dashed lines represent negative values. The vertical dashed and dotted lines mark $z=\hd$ and $z=\zsd=4\hd$, respectively.} 
\label{fig:eigen_func_fall}
\end{figure*}

Figure \ref{fig:flow_struct_fall} shows the gas and dust density perturbations and their streamlines for $\qthree=5,\;\taus=0.3\;\tilde{D}_x=3\times10^{-4},\;\tilde{D}_z=1\times10^{-4},\;\epsilon(0)=8$ and $kH=14$. Figure \ref{fig:eigen_func_fall} shows each perturbed values as a function of $z$ as in Figure \ref{fig:eigen_func_fiducial}. The dust- and gas-flow structures at $z\gtrsim0.02H$ are similar to the flow structure in the case 1 while the lower layer shows different flow structures. 

\begin{figure}[t!]%[htp] or [H]
	\begin{center}
	\hspace{0pt}\raisebox{20pt}{
	\includegraphics[width=0.5\columnwidth]{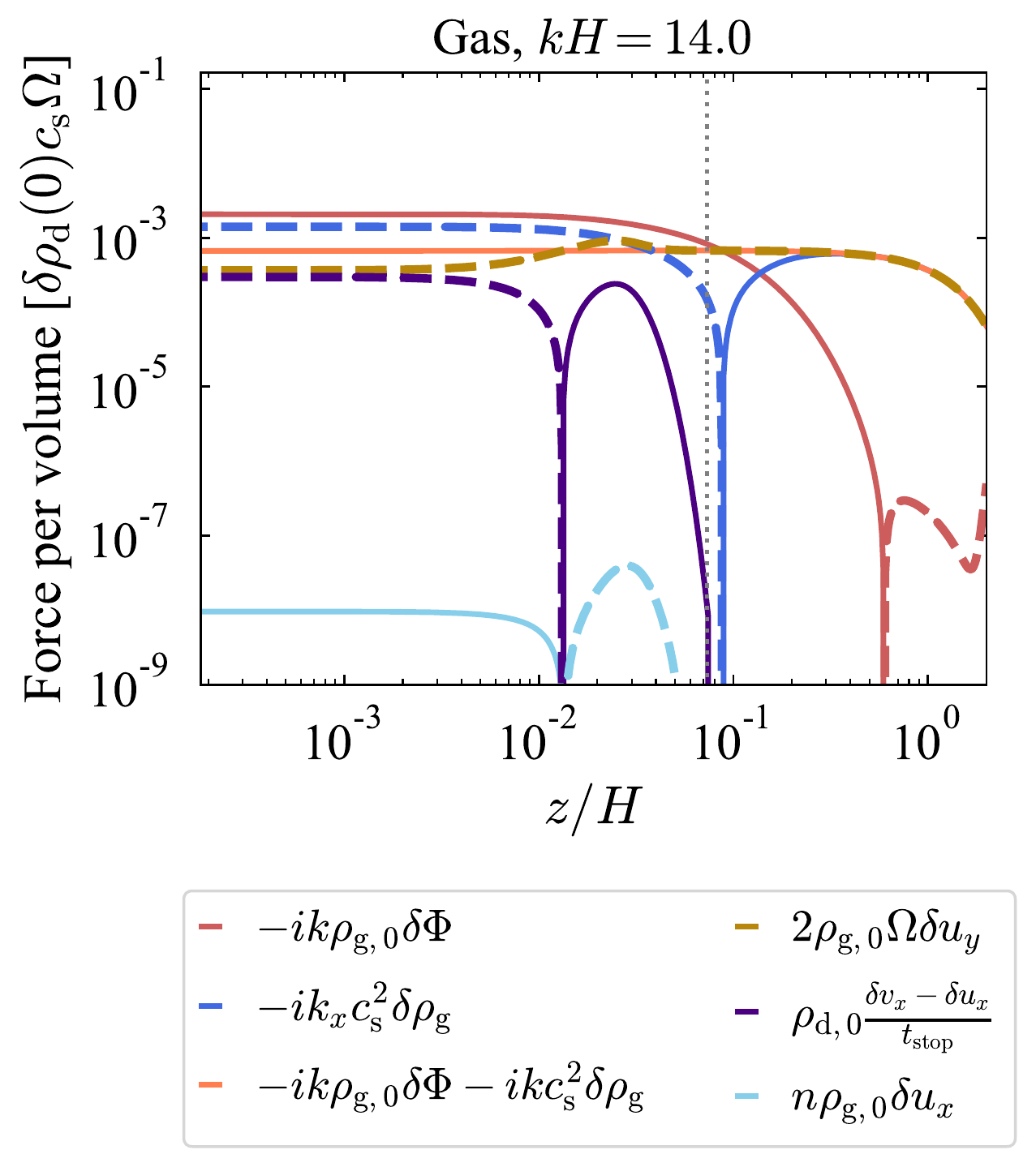} %2 column
%	\hspace{0pt}\raisebox{20pt}{
%	\includegraphics[width=0.5\columnwidth]{./Q3d5_alpr3e-4_alpz1e-4_taus03_dgmid8_Hs2Hg_gas_radial_Force2.pdf} %1 column
	}
	\end{center}
	\vspace{-30pt}
\caption{The radial force on gas and its balance at $x=-\lambda/4$ for the dust-fall mode. As in Figure \ref{fig:eigen_func_fall}, a solid (dashed) line represents a region where a physical value is positive (negative). The vertical gray dotted line marks $z=\zsd=4\hd$.} 
\label{fig:gas_radial_force_fall_mode}
\end{figure}

First, the gas is much more incompressible and shows much smaller $\delta\rhog$ than in the cases of the main part of this paper. Besides, the gas flow is divided into two parts. One is the flow around the midplane, where the gas circulation is counterclockwise (clockwise) at $x<0$ ($x>0$). The other is the meridional flow in the relatively upper region, which is similar to the flow structure in the case 1 (Figure \ref{fig:flow_struct_fiducial}). We find that the gas motion toward the density maximum around the midplane is driven by the self-gravity, which can bee seen in Figure \ref{fig:gas_radial_force_fall_mode}.

The dust shows meridional flow around the midplane while the upper dust moves almost radially. Although the meridional dust flow is also found in the case 4 (Figure \ref{fig:flow_struct_2nd_type}), the dust-fall mode shows the opposite direction of the circulation, e.g., clockwise motion at $x<0$. The dust-density enhancement is driven by the fall flow that is launched at the negative-$\delta\sigmad$ region (e.g., $x=-0.2H$ in Figure \ref{fig:flow_struct_fall}), passes in the radial direction above the midplane, and falls toward the positive-$\delta\sigmad$ region (e.g., $x=0$). The vertical mass flux of dust is larger in magnitude than the vertical gas flow (the right bottom panel of Figure \ref{fig:eigen_func_fall}), which is in contrast to the case 1. We can also see the effect of the vertical flow in the dust density perturbation profile (the left top panel). There is a negative-$\delta\rhod$ region at $z>\hd$, which is due to the efficient vertical flow of dust.

We observe the dust-fall flow when we adopt weaker vertical diffusion than radial diffusion. The double meridional gas flow with the dust-fall flow appears when the dust-to-gas ratio is much larger as in the case of Figure \ref{fig:flow_struct_fall} ($\epsilon(0)=8$). If $\epsilon(0)$ is much larger than unity, the small-scale dust clumping is expected to proceed efficiently as seen in numerical simulations of streaming instability \citep[e.g.,][]{Schreiber2018,Schaffer2021}. Since the dust-clumping reduces the dust diffusion efficiency as found in \citet{Schreiber2018}, the assumption of the constant $\tilde{D}_z$ and $\tilde{D}_r$ in the present analyses would be invalid. Numerical simulations will be necessary to figure out whether or not the dust-fall-type secular GI appears in more realistic situations.

\section{List of parameters}\label{app:params_cond}
Table \ref{tab:params_appendix} summarizes sets of parameters used to plot Figure \ref{fig:growth_condition}.

\begin{deluxetable}{cccccccc}[ht]
\tablecaption{Sets of parameters} \label{tab:params_appendix}
\tablehead{
\colhead{$\qthree$} & 
\colhead{$\taus$} & 
\colhead{$\tilde{D_r}$} &
\colhead{$\tilde{D_z}$} & 
\colhead{$\epsilon(0)$} &
\colhead{$\tilde{D_z}/\taus$} &
\colhead{$Q_{\dst}\sqrt{\varepsilon}$} &
\colhead{unstable? (Y/N)}  
}
\startdata
2.0 &
0.5 &
1.0$\times 10^{-4}$ &
1.0$\times 10^{-3}$ &
1.0 &
2.0$\times 10^{-3}$ &
0.278 &
 Y \\ 
2.0 &
0.2 &
1.0$\times 10^{-4}$ &
1.0$\times 10^{-3}$ &
1.0 &
5.0$\times 10^{-3}$ &
0.351 &
 Y \\ 
2.0 &
0.2 &
1.0$\times 10^{-4}$ &
1.0$\times 10^{-3}$ &
0.2 &
5.0$\times 10^{-3}$ &
0.744 &
 Y \\ 
2.0 &
0.2 &
1.0$\times 10^{-4}$ &
2.0$\times 10^{-5}$ &
1.0 &
1.0$\times 10^{-4}$ &
0.926 &
 Y \\ 
2.0 &
0.5 &
1.0$\times 10^{-4}$ &
3.0$\times 10^{-4}$ &
0.3 &
6.0$\times 10^{-4}$ &
0.656 &
 Y \\ 
2.0 &
0.5 &
1.0$\times 10^{-4}$ &
1.0$\times 10^{-3}$ &
0.05  &
2.0$\times 10^{-3}$ &
1.169 &
 N \\ 
2.0 &
0.5 &
1.0$\times 10^{-4}$ &
3.0$\times 10^{-4}$ &
0.1  &
6.0$\times 10^{-4}$ &
1.120 &
 N \\ 
2.0 &
0.5 &
1.0$\times 10^{-4}$ &
3.0$\times 10^{-4}$ &
0.5 &
6.0$\times 10^{-4}$ &
0.515 &
 Y \\ 
2.0 &
0.5 &
1.0$\times 10^{-4}$ &
1.0$\times 10^{-3}$ &
0.2 &
2.0$\times 10^{-3}$ &
0.591 &
 Y \\ 
2.0 &
0.5 &
1.0$\times 10^{-4}$ &
1.0$\times 10^{-4}$ &
1.0 &
2.0$\times 10^{-4}$ &
0.493 &
 Y \\ 
2.0 &
0.5 &
1.0$\times 10^{-4}$ &
1.0$\times 10^{-3}$ &
0.5 &
2.0$\times 10^{-3}$ &
0.382 &
 Y \\ 
2.0 &
0.2 &
1.0$\times 10^{-4}$ &
2.0$\times 10^{-5}$ &
2.0 &
1.0$\times 10^{-4}$ &
0.688 &
 Y \\ 
2.0 &
0.2 &
1.0$\times 10^{-4}$ &
1.0$\times 10^{-3}$ &
0.1  &
5.0$\times 10^{-3}$ &
1.044 &
 N \\ 
2.0 &
0.2 &
1.0$\times 10^{-4}$ &
2.0$\times 10^{-5}$ &
0.8  &
1.0$\times 10^{-4}$ &
1.024 &
 N \\ 
2.0 &
0.5 &
3.0$\times 10^{-4}$ &
3.0$\times 10^{-4}$ &
0.8 &
6.0$\times 10^{-4}$ &
0.718 &
 Y \\ 
2.0 &
0.5 &
1.0$\times 10^{-4}$ &
3.0$\times 10^{-4}$ &
0.2 &
6.0$\times 10^{-4}$ &
0.798 &
 Y \\ 
2.0 &
0.5 &
1.0$\times 10^{-4}$ &
3.0$\times 10^{-5}$ &
0.3  &
6.0$\times 10^{-5}$ &
1.165 &
 N \\ 
2.0 &
0.2 &
1.0$\times 10^{-4}$ &
1.0$\times 10^{-3}$ &
0.3 &
5.0$\times 10^{-3}$ &
0.612 &
 Y \\ 
2.0 &
0.5 &
1.0$\times 10^{-4}$ &
1.0$\times 10^{-3}$ &
0.3 &
2.0$\times 10^{-3}$ &
0.486 &
 Y \\ 
2.0 &
0.5 &
1.0$\times 10^{-4}$ &
3.0$\times 10^{-5}$ &
1.0 &
6.0$\times 10^{-5}$ &
0.665 &
 Y \\ 
2.0 &
0.2 &
1.0$\times 10^{-4}$ &
1.0$\times 10^{-3}$ &
0.5 &
5.0$\times 10^{-3}$ &
0.481 &
 Y \\ 
2.0 &
0.5 &
1.0$\times 10^{-4}$ &
1.0$\times 10^{-3}$ &
0.1 &
2.0$\times 10^{-3}$ &
0.830 &
 Y \\ 
2.0 &
0.5 &
1.0$\times 10^{-4}$ &
3.0$\times 10^{-4}$ &
1.0 &
6.0$\times 10^{-4}$ &
0.375 &
 Y \\ 
2.0 &
0.5 &
1.0$\times 10^{-4}$ &
3.0$\times 10^{-5}$ &
0.5 &
6.0$\times 10^{-5}$ &
0.914 &
 Y \\ 
2.0 &
0.5 &
1.0$\times 10^{-4}$ &
1.0$\times 10^{-4}$ &
0.2  &
2.0$\times 10^{-4}$ &
1.050 &
 N \\ 
2.0 &
0.5 &
1.0$\times 10^{-4}$ &
1.0$\times 10^{-3}$ &
0.08 &
2.0$\times 10^{-3}$ &
0.926 &
 N \\ 
2.0 &
0.5 &
3.0$\times 10^{-4}$ &
3.0$\times 10^{-4}$ &
1.0 &
6.0$\times 10^{-4}$ &
0.649 &
 Y \\ 
2.0 &
0.5 &
1.0$\times 10^{-4}$ &
1.0$\times 10^{-4}$ &
0.3 &
2.0$\times 10^{-4}$ &
0.863 &
 Y \\ 
2.0 &
0.2 &
1.0$\times 10^{-4}$ &
1.0$\times 10^{-3}$ &
0.14  &
5.0$\times 10^{-3}$ &
0.885 &
 N \\ 
2.0 &
0.5 &
3.0$\times 10^{-4}$ &
3.0$\times 10^{-4}$ &
0.3  &
6.0$\times 10^{-4}$ &
1.136 &
 N \\ 
2.0 &
0.5 &
1.0$\times 10^{-4}$ &
1.0$\times 10^{-4}$ &
0.5 &
2.0$\times 10^{-4}$ &
0.677 &
 Y \\ 
2.0 &
0.5 &
3.0$\times 10^{-4}$ &
3.0$\times 10^{-4}$ &
0.5 &
6.0$\times 10^{-4}$ &
0.891 &
 Y \\ 
3.0 &
1.0 &
1.0$\times 10^{-4}$ &
1.0$\times 10^{-4}$ &
0.3 &
1.0$\times 10^{-4}$ &
1.028 &
 N \\ 
3.0 &
0.5 &
1.0$\times 10^{-4}$ &
1.0$\times 10^{-4}$ &
0.5 &
2.0$\times 10^{-4}$ &
0.956 &
 N \\ 
3.0 &
0.5 &
1.0$\times 10^{-4}$ &
1.0$\times 10^{-4}$ &
1.0 &
2.0$\times 10^{-4}$ &
0.692 &
 Y \\ 
3.0 &
0.2 &
1.0$\times 10^{-4}$ &
1.0$\times 10^{-4}$ &
1.0 &
5.0$\times 10^{-4}$ &
0.871 &
 Y \\ 
3.0 &
1.0 &
1.0$\times 10^{-4}$ &
1.0$\times 10^{-4}$ &
0.5 &
1.0$\times 10^{-4}$ &
0.804 &
 Y \\ 
3.0 &
0.2 &
1.0$\times 10^{-4}$ &
1.0$\times 10^{-4}$ &
2.0 &
5.0$\times 10^{-4}$ &
0.641 &
 Y \\ 
3.0 &
0.2 &
1.0$\times 10^{-4}$ &
1.0$\times 10^{-4}$ &
0.8 &
5.0$\times 10^{-4}$ &
0.965 &
 N \\ 
3.0 &
1.0 &
1.0$\times 10^{-4}$ &
1.0$\times 10^{-4}$ &
0.8 &
1.0$\times 10^{-4}$ &
0.644 &
 Y \\ 
3.0 &
0.5 &
1.0$\times 10^{-4}$ &
1.0$\times 10^{-4}$ &
0.8 &
2.0$\times 10^{-4}$ &
0.767 &
 Y \\ 
5.0 &
0.3 &
3.0$\times 10^{-4}$ &
3.0$\times 10^{-4}$ &
10.0 &
1.0$\times 10^{-3}$ &
0.624 &
 Y \\ 
5.0 &
0.3 &
3.0$\times 10^{-4}$ &
1.0$\times 10^{-4}$ &
8.0 &
3.3$\times 10^{-4}$ &
0.884 &
 Y \\ 
5.0 &
0.3 &
3.0$\times 10^{-4}$ &
1.0$\times 10^{-4}$ &
5.0  &
3.3$\times 10^{-4}$ &
1.057 &
 N \\ 
5.0 &
0.3 &
3.0$\times 10^{-4}$ &
1.0$\times 10^{-4}$ &
7.0 &
3.3$\times 10^{-4}$ &
0.929 &
 Y \\ 
5.0 &
0.3 &
3.0$\times 10^{-4}$ &
3.0$\times 10^{-4}$ &
5.0 &
1.0$\times 10^{-3}$ &
0.806 &
 Y \\ 
5.0 &
0.3 &
3.0$\times 10^{-4}$ &
1.0$\times 10^{-4}$ &
10.0 &
3.3$\times 10^{-4}$ &
0.816 &
 Y \\ 
5.0 &
0.3 &
3.0$\times 10^{-4}$ &
3.0$\times 10^{-4}$ &
3.0  &
1.0$\times 10^{-3}$ &
0.993 &
 N \\ 
\enddata
\end{deluxetable}

\bibliographystyle{aasjournal}
\bibliography{rttominaga}

%\begin{thebibliography}{}
%
%\bibitem[Astropy Collaboration et al.(2013)]{2013A&A...558A..33A} Astropy Collaboration, Robitaille, T.~P., Tollerud, E.~J., et al.\ 2013, \aap, 558, A33 
%\bibitem[Bertin \& Arnouts(1996)]{1996A&AS..117..393B} Bertin, E., \& Arnouts, S.\ 1996, \aaps, 117, 393 
%\bibitem[Corrales(2015)]{2015ApJ...805...23C} Corrales, L.\ 2015, \apj, 805, 23
%\bibitem[Ferland et al.(2013)]{2013RMxAA..49..137F} Ferland, G.~J., Porter, R.~L., van Hoof, P.~A.~M., et al.\ 2013, \rmxaa, 49, 137
%\bibitem[Hanisch \& Biemesderfer(1989)]{1989BAAS...21..780H} Hanisch, R.~J., \& Biemesderfer, C.~D.\ 1989, \baas, 21, 780 
%\bibitem[Lamport(1994)]{lamport94} Lamport, L. 1994, LaTeX: A Document Preparation System, 2nd Edition (Boston, Addison-Wesley Professional)
%\bibitem[Schwarz et al.(2011)]{2011ApJS..197...31S} Schwarz, G.~J., Ness, J.-U., Osborne, J.~P., et al.\ 2011, \apjs, 197, 31  
%\bibitem[Vogt et al.(2014)]{2014ApJ...793..127V} Vogt, F.~P.~A., Dopita, M.~A., Kewley, L.~J., et al.\ 2014, \apj, 793, 127  
%
%\end{thebibliography}


\end{document}